\documentclass[twocolumn]{aastex631}
\newcommand{\HI}{\hbox{{\rm H}\kern 0.2em{\sc i}}}
\usepackage{CJK}
\accepted{April 18, 2025}
\shorttitle{Stars and {\HI} gas in Virgo filaments and groups}
\shortauthors{Yoon et al.}
\begin{document}
\begin{CJK*}{UTF8}{mj}
%\begin{CJK*}{UTF8}{gbsn} Note: doesn't work for KOREAN (author's name)

%%%%%%%%%%%%%%%%%%%%%%%%%%%%%%%%%%% Title 
\title{Mapping the Star Formation and {\HI} Gas Properties of Galaxies Along Large-scale Structures\\Around the Virgo Cluster}
%%%%%%%%%%%%%%%%%%%%%%%%%%%%%%%%%%% 

%%%%%%%%%%%%%%%%%%%%%%%%%%%%%%%%%%% Authors
%\correspondingauthor{Aeree Chung}
\email{hiyoon.astro@gmail.com, achung@yonsei.ac.kr}
%NOTE: The corresponding author is Aeree Chung. We prefer to include email addresses only within the manuscript.

\author[0000-0003-4048-2203]{Hyein Yoon (윤혜인)}
\affiliation{Department of Astronomy, Yonsei University, 50 Yonsei-ro, Seodaemun-gu, Seoul 03722, Republic of Korea}
\affiliation{Institute for Data Innovation in Science, Seoul National University, 1 Gwanak-ro, Gwanak-gu, Seoul 08826, Republic of Korea}
\affiliation{Astronomy Program, Department of Physics and Astronomy, Seoul National University, 1 Gwanak-ro, Gwanak-gu, Seoul 08826, Republic of Korea}
\affiliation{Sydney Institute for Astronomy, School of Physics A28, University of Sydney, NSW 2006, Australia}

\author[0000-0003-4264-3509]{O. Ivy Wong}
\affiliation{CSIRO Astronomy \& Space Science, PO Box 1130, Bentley, WA 6102, Australia}
\affiliation{ICRAR-M468, University of Western Australia, Crawley, WA 6009, Australia}

\author[0000-0003-1440-8552]{Aeree Chung}
\affiliation{Department of Astronomy, Yonsei University, 50 Yonsei-ro, Seodaemun-gu, Seoul 03722, Republic of Korea}

\author{Shan Huang}
\affiliation{Center for Cosmology and Particle Physics, New York University, New York, NY 10003, USA}
%%%%%%%%%%%%%%%%%%%%%%%%%%%%%%%%%%%

%%%%%%%%%%%%%%%%%%%%%%%%%%%%%%%%%%% Abstract (250 word-limit)
\begin{abstract}
We investigate the star formation and neutral atomic hydrogen ({\HI}) gas properties of galaxies along three large-scale filaments and two galaxy groups in the wide field around the Virgo cluster. Our goal is to understand how galaxies are processed in low-density environments before falling into high-density regions. Combining the spatial distribution of galaxies with multiwavelength colors such as W3$-$W1, NUV$-r$, and \textit{g$-$r}, we find a predominance of blue galaxies across the structures, indicating normal-to-enhanced star formation, similar to that of isolated galaxies. However, one filament and one group show a significant number of red galaxies (32\% and 20\%, respectively), suggesting that star formation has been suppressed in low-density environments before reaching high-density regions. Intriguingly, these red galaxies span a wide range of stellar masses, and the presence of red dwarfs support that not only mass but also environment plays an important role in the quenching of star formation in cluster outskirts. One particular filament, potentially connected to Virgo, already has a group of red populations outside Virgo's $R_{200}$, making these galaxies good candidates for being ``\textit{preprocessed}'' before entering the Virgo cluster. In addition, several galaxies in the filaments and groups possess relatively low {\HI} gas contents, similar to cluster galaxies. However, the overall fraction of {\HI}-deficient galaxies is not as significantly high as the fraction of red galaxies in these structures. This suggests that {\HI} gas properties are less influenced by the environment than star formation properties in low-density regions, possibly due to gas replenishment through accretion.
\end{abstract}
%%%%%%%%%%%%%%%%%%%%%%%%%%%%%%%%%%%

%%%%%%%%%%%%%%%%%%%%%%%%%%%%%%%%%%% Keywords
% https://astrothesaurus.org
%%%%%%%%%%%%%%%%%%%%%%%%%%%%%%%%%%%

%%%%%%%%%%%%%%%%%%%%%%%%%%%%%%%%%%% Section 1. Introduction
\section{Introduction}
\label{sec1:intro}

While moving from a sparse to a heavily dense environment, a large fraction of spiral galaxies may undergo transformation to S0 or elliptical galaxies \citep[][]{dressler80,dressler97} featuring a quenched star formation. The suppression of star formation in galaxies is a complex process, and the exact mechanisms driving this transformation remain an open question. However, numerous studies have shown that the environment plays a crucial role in this process \citep[e.g.,][]{balogh99,kauffmann04,boselli06,poggianti06,peng10,scoville13}. In high-density regions such as clusters, the quenching of star formation is strongly dependent upon various physical processes, including gas stripping. Even before falling to clusters, galaxies can also be affected by their surroundings in low-density environments  \citep[e.g.,][]{odekon18,kleiner21,vulcani21}. A key aspect of this precluster evolution is ``\textit{preprocessing}'', where galaxies undergo environmental transformation within low-density regions, such as groups, subclusters, and filaments. Recent observations and simulations have suggested this process \citep[e.g.,][]{fujita04,cortese06,mahajan13}, but further evidence is needed to fully understand its impact on galaxy evolution.

Neutral atomic hydrogen ({\HI}) gas is one of the primary components of galaxies affected by environmental effects. For instance, \citet{solanes01} statistically show that a great number of {\HI}-poor galaxies are present mostly in the high-density cluster environment, yet still a good fraction of galaxies is found outside of clusters. In the outskirts of galaxy clusters, galaxies can also possess truncated {\HI} gas disks, indicating past gas stripping \citep[e.g.,][]{chung09,yoon17}. \citet{hess13} show that the loss of the {\HI} gas can happen in group environments based on the spatial distribution of the gas content of galaxy groups in different redshift ranges. \citet{janowiecki17} show the higher {\HI} gas content and enhanced star formation in low-mass group central galaxies, suggesting fueling from the cosmic web or minor mergers. In fact, there are galaxies, possibly moving along the filament, accreting gas from the intergalactic medium, as shown by {\HI} imaging studies \citep[e.g.,][]{oosterloo07,sancisi08,deblok14}. These ongoing efforts to uncover the mechanisms driving {\HI} depletion or accretion in low-density regions are valuable but remain largely constrained to the local Universe. More direct observational evidence will be a key scientific focus for next-generation high-sensitivity telescopes.
 
Combining {\HI} with star formation properties of galaxies offers more complete understanding of the physical mechanism shaping galactic properties in low-density environments. \citet{huang12} examined UV, optical, and {\HI} data, suggesting that environmental effects likely play significant roles in regulating and quenching star formation in low-mass galaxies. \citet{cybulski14} show that a large number of galaxies in groups have a strong dependence on the environments. \citet{jaffe16} combine {\HI}, IR, optical, UV, and X-ray data and present that many passive galaxies, which are located outside the core of the cluster at $z=0.2$. \citet{luber19} find that redder and more massive galaxies are closer to the cosmic web, based on their analysis of the COSMOS field at redshifts from 0 to 0.4. \citet{reynolds22} investigate the effect of the environment by comparing measured {\HI} and star formation properties in populations of cluster, infall, and field galaxies. Notably, a rapid decrease in the {\HI}-detected fraction of infalling galaxies is found at a projected distance of $\sim$1.5$R_{200}$. These studies highlight that stellar components in galaxies act as effective tracers of preprocessing alongside the observed signatures in {\HI} gas disks.

Recent theoretical studies provide a general perspective of the environmental impacts on the outskirts of clusters, particularly within the cosmic web. \citet{tonnesen07} indicate that the gas stripping of galaxies can take place far outside the cluster center. Cosmological simulations by \citet{vijayaraghavan13} present that the significant impact of the group environment on galaxies, caused by enhanced tidal interactions and ram pressure stripping, prior to cluster infall. \citet{jung18} show that preprocessing preferentially affects low-mass satellite galaxies, leading them to become gas-poor in group-sized halos before they enter the cluster. \citet{stevens19} find that the {\HI} deficit of a satellite galaxy depends on its time since infall in the same way as its star formation rate (SFR). \citet{galarraga20} show that the state of gas within filaments depends on the presence of halos and the large-scale environment. \citet{malvasi22} present star formation that shows the largest variation with distance from the cosmic web features compared to the relatively smaller variation with mass. It also shows the strongest relation to the local environment of galaxies. The theoretical results can synergize with observational data, complementing each other in addressing resolution limitations, multiphase gas behavior, and distinguishing various environmental effects.

Previous and ongoing studies of {\HI} gas, star formation, and theoretical models have advanced our understanding of preprocessing; however, they are primarily focused on statistical analyses. A more detailed, close-up investigation is still crucial for a comprehensive view of galaxy evolution in low-density environments. Virgo and the surrounding large-scale structures are ideal places to study galaxies spanning diverse environments. There have been several recent studies focusing on Virgo filaments. \citet{kim16} identified eight large-scale structures in the volume surrounding the Virgo cluster, and \citet{lee21} show negative color and stellar mass gradients of galaxies in filaments, suggesting mass assembly depending on the efficiency of galaxy mergers at different distances from the filament spine. \citet{chung21} investigate the chemical properties of star-forming dwarf galaxies in Virgo filaments and show that, at a given stellar mass, these galaxies exhibit lower metallicity and higher specific star formation rates (sSFR) compared to those in the Virgo cluster on average. \citet{castignani22a} present that a large fraction of filament galaxies around Virgo are deficient either in {\HI} or H$_{2}$, or both. They extend their work to the cosmic web in the wide field around Virgo, reaching up to 12~virial radii \citep{castignani22b}. \citet{zakharova24} used semianalytic models to reproduce the atomic and molecular gas properties in filaments. The results confirm a decrease in {\HI}-deficient galaxies from cluster to filaments, supporting the observed trend in  \citet{castignani22a,castignani22b} that filaments exhibit intermediate properties between cluster and field galaxies.

In this paper, we aim to investigate the role of large-scale filaments and group environments in shaping galaxy evolution before cluster infall. Specifically, we analyze the spatial distribution of galaxies in these structures alongside their star formation properties, traced through multiwavelength colors. Furthermore, we compare the {\HI} gas content of galaxies across different density environments to assess how physical processes influence gas retention and depletion in low-density regions. Through this approach, we seek to provide new insights into the mechanisms driving galaxy evolution within the cosmic web before they reach the cluster environment.

This paper is organized as follows. Section~\ref{sec2:sample} introduces our sample selection criteria and multiwavelength data used. Section~\ref{sec3:sf_indicators} describes the color-based star formation indicators. In Section~\ref{sec4:sf_environments}, we combine the spatial distribution of the galaxies with the star formation properties. The {\HI} gas content of the sample is studied in Section~\ref{sec5:hi_environments}. In Section~\ref{sec6:discussion}, we discuss the preprocessing of the galaxies in low-density environments. Finally, in Section~\ref{sec7:summary}, we summarize and conclude.

Throughout this study, we adopt $\Lambda$CDM cosmological parameters, $\Omega_{\rm M}=0.30$, $\Omega_{\Lambda}=0.70$, and $H_{0}=70$~km~s$^{-1}$~Mpc$^{-1}$. The sample galaxies in this work are selected from various large-scale structures, each within a carefully selected specific radial velocity range. The majority of interpretations rely on color-based analysis within the same radial velocity plane, which is minimally affected by distance information. Luminosity distance is used solely to obtain more accurate stellar mass estimates for each galaxy, and the distances to our sample have been estimated by using \verb'CosmoloPy' Python package \footnote{\href{http://roban.github.io/CosmoloPy/}{http://roban.github.io/CosmoloPy/}}. Note that Virgocentric flow and other peculiar velocities are not particularly considered in this work.
%%%%%%%%%%%%%%%%%%%%%%%%%%%%%%%%%%%

%%%%%%%%%%%%%%%%%%%%%%%%%%%%%%%%%%% Seciton 2. Sample
\section{Sample and Data}
\label{sec2:sample}

%%%%%%%%%%%%%%%%%%%%%%%%%%%%%%%%%%% Figure 1 (column width)
\begin{figure}
\epsscale{1.2}
\plotone{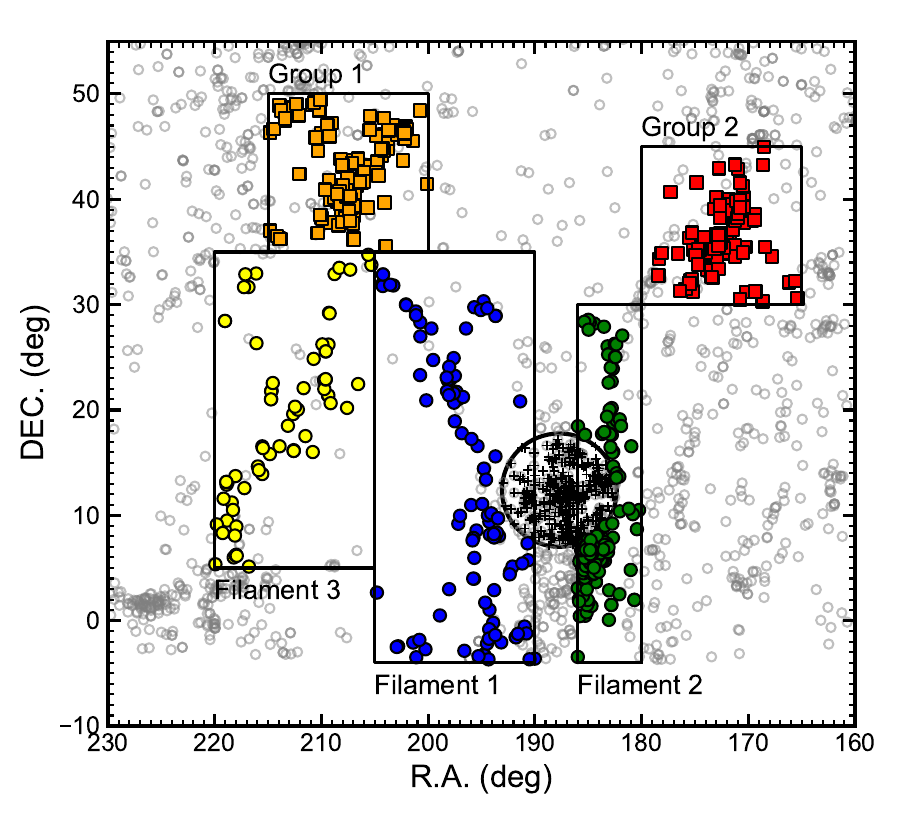}
\caption{Spatial distribution of the galaxies in three large-scale filaments and two galaxy groups around Virgo. Black boxes represent the sky coverage of each structure. Blue is Filament~1, green is Filament~2, yellow is Filament~3, orange is Group~1, and red is Group~2. Background gray circles are the galaxies which cover the total radial velocity range of our sample, $1300 < cz < 3200$~km~s$^{-1}$. Black crosses in the middle are Virgo reference, and they are enclosed by the solid circle which corresponds to $R_{200}$ of Virgo \citep[$\sim$5.4 deg;][]{mclaughlin99,ferrarese12}. The data is taken from the SDSS DR12 database. From Filaments~1 and 2, an overlapped sample with Virgo reference has been carefully excluded.
\label{fig:f1_sample}}
\end{figure} 
%%%%%%%%%%%%%%%%%%%%%%%%%%%%%%%%%%%

%%%%%%%%%%%%%%%%%%%%%%%%%%%%%%%%%%% Figure 2 (column width)
\begin{figure}
\epsscale{1.2}
\plotone{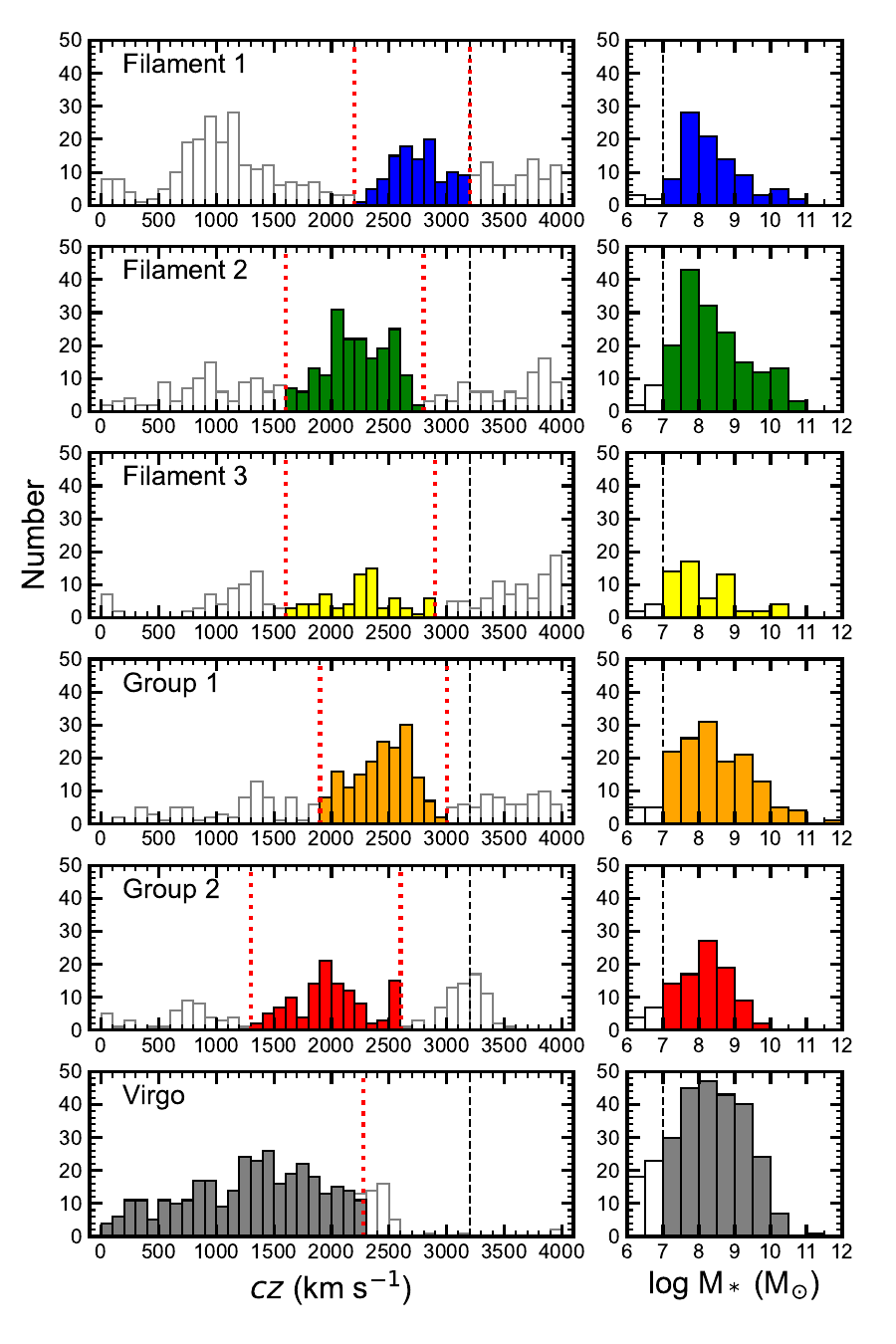}
\caption{Left: Radial velocity distribution of the sample. The color scheme is the same as shown in Figure~\ref{fig:f1_sample}. The dotted red lines indicate the selected velocity ranges of individual structures. The black dashed line shows the upper velocity limit of ALFALFA $\alpha$.100 catalog ($cz=3200$~km~s$^{-1}$). Background open bars show the velocity distribution of all galaxies in each region with $cz < 4000$~km~s$^{-1}$. Right: Stellar mass distribution of the galaxies. The black dashed line indicates the lower stellar mass cut adopted in this work (log~$M_{*} = 7$). The data is from the SDSS DR12 database. 
\label{fig:f2_histo}}
\end{figure}
%%%%%%%%%%%%%%%%%%%%%%%%%%%%%%%%%%%%%%%%%

%%%%%%%%%%%%%%%%%%%%%%%%%%%%%%%%%%% Table 1 (page width)
\begin{deluxetable*}{lcccccccc}
\tablecaption{General Properties of Five Large-scale Structures and Virgo Sample
\label{tab1:sample}}
\setlength{\tabcolsep}{7pt} 
\tablehead{
\colhead{Structure} & \colhead{R.A. range} & \colhead{Decl. range} & \colhead{$cz$ Range} & \colhead{Num of Gal.} & \colhead{Num of Gal.} & \colhead{Num of Gal.} & \colhead{Num of Gal.} & \colhead{Ref.} \\
\colhead{} & \colhead{} & \colhead{} & \colhead{} & (Optical) & (IR) & (UV) & ({\HI}) & \\
\colhead{} & (deg) & (deg) & (km~s$^{-1}$) & \colhead{} & \colhead{} & \colhead{} & \colhead{} & \colhead{}
}
\decimalcolnumbers
\startdata
Filament~1 & [190 : 205]  & [$-4$ : 35] & [2200 : 3200] & 86 & 37 & 77 & 45 & (a) \\
Filament~2 & [180 : 186]  & [$-4$ : 30] & [1600 : 2800] & 153 & 72 & 142 & 76 & (a) \\
Filament~3 & [205 : 220]  & [5 : 35] & [1600 : 2900] & 55 & 28 & 42 & 45 & \\
\hline
Group~1 & [200 : 215]  & [35 : 50] & [1900 : 3000] & 137 & 78 & 125 & 2 (44\tablenotemark{\scriptsize $a$}) &  (b) \\
Group~2 & [165 : 180]  & [30 : 45] & [1300 : 2600] & 79 & 36 & 62 & 33 (52\tablenotemark{\scriptsize $a$}) &\\
\hline
Virgo & \multicolumn{2}{c}{$d_{\rm M87} < R_{200}$ ($\sim$5.4 deg)} & [$-98$ : 2274]\tablenotemark{\scriptsize b} & 222 & 109 & 194 & 69 & (c)\\
\enddata
\tablecomments{~Column~(1): Name of the structure. Columns~(2), (3), and (4): R.A., decl., and radial velocity ranges. Columns~(5), (6), (7), and (8): Number of galaxies detected in optical, IR, UV and {\HI}, respectively. Column~(9): Other references which studied each structure previously: (a) \citet{kim16}; (b) \citet{tully08}; and (c) \citet{mei07}.\\
\tablenotemark{\scriptsize a}The {\HI} detections in these groups are added from HyperLeda \citep[][]{paturel03}.\\
\tablenotemark{\scriptsize b}This range covers two-dispersion deviation ($2\sigma_{\rm Virgo}=1186$~km~s$^{-1}$) from the mean radial velocity of the M87 group \citep[$v=1088$~km~s$^{-1}$;][]{mei07}}
\end{deluxetable*}
%%%%%%%%%%%%%%%%%%%%%%%%%%%%%%%%%%%

%%%%%% subsection 2.1
\subsection{Large-scale Filaments and Groups Around the Virgo Region}

This work makes use of galaxy colors and gas content derived from UV, optical, IR, and {\HI} data. Further details on the datasets are provided in the following subsection. To identify large-scale structures around the Virgo cluster, we primarily use optical data from the Sloan Digital Sky Survey Data Release 12 \citep[SDSS DR12;][]{sdssdr12}. We define our working sample by applying a radial velocity cut of $1300 < cz < 3200$~km~s$^{-1}$, which encompasses the core velocity range of the Virgo cluster as well as its surrounding large-scale structures. This velocity limit ensures that we capture both the cluster and the extended foreground and background structures potentially associated with it. Using this dataset, we have selected three filaments (Filaments~1, 2, and 3) and two overdense regions (Groups~1 and 2) within narrow windows in both sky position and radial velocity, based on the observed distribution of galaxies. 

Figure~\ref{fig:f1_sample} shows the sky distribution of the five selected structures and their member galaxies. For comparison, reference galaxies in the central Virgo region ($<R_{200}$) are also included, but these do not overlap with Filaments~1 and 2. The detailed criteria for spatial and radial velocity selections, along with the number of member galaxies in each structure, are summarized in Table~\ref{tab1:sample}. In Figure~\ref{fig:f2_histo}, the distribution of radial velocity (left) and stellar mass (right) of the galaxies are shown. The radial velocity distribution has been fitted by Gaussian, and a 2$\sigma$ deviation from the mean is roughly used to choose the total velocity range of the structures. Filament~3 is notably identifiable as a clear filament, although the member galaxies do not show a Gaussian distribution. In this case, we carried out a visual selection, making a careful cut considering a combination of their spatial and velocity distributions. 

There have been various refined methods for identifying cosmic webs and groups until date, based on the persistent structures identification \citep[DisPerSE;][]{sousbie11}, Friends-of-Friends algorithm \citep[FoF;][]{huchra82}, and Minimum Spanning Tree \citep[MST;][]{barrow85} algorithm. Here, it is important to clarify that our primary focus in this work is not on precisely defining these structures or identifying member galaxies, given the complexity of the local Universe. We aim to minimize the removal of foreground and background galaxies from each structure while investigating trends across the wide field around Virgo. Nonetheless, we also conducted various tests to validate our sample selection (see Appendix~A). Our main objective is to making a wide-field map encompassing the selected structures in order to observe in projection any changes in galactic properties within the context of the large-scale environment. In this context, all the structures we have selected are excellent representatives of low-density environments, and they may not necessarily be connected to the Virgo cluster. The Virgo region is nearby and provides a wealth of detailed information; however, it is a complex and unvirialized local structure that cannot be entirely explained by the Hubble expansion alone due to its nearness. This complexity implies potential uncertainties in accurately identifying large-scale structures using any developed identification methods. Thus, we tried to keep potential peripheral galaxies around the filaments and groups while minimizing the exclusion of outliers. In Section~\ref{sec6:discussion}, we verify our wide-field results with those based on the Voronoi Tessellation and MST quantification of projected large-scale structures. Our sample selection, based on the sky and velocity distributions, aligns well with the results obtained from the established tools.

The large-scale structures around the Virgo cluster have been referenced by other studies \citep{tully82,kim16,castignani22b}. \citet{kim16} analyzed Filament~1 and 2, the most noticeable filaments in the northern sky of Virgo. Group~1 (NGC~5353/4 group) was studied by \citet{tully08}. In this work, we also select Filament~3, which is quite well defined but relatively not well studied. Group~1 is physically connected to Filament~1, whereas it remains unclear if Group~2 is associated with Filament~2. However, this overdense group is located in a similar range of radial velocity as Filament~2.

In addition to the galaxies within the five large-scale structures, we also select reference galaxies in the central region of the Virgo cluster. They are located within $R_{200}$ of Virgo \citep[$\sim$5.4 deg;][]{ferrarese12,mclaughlin99}, and the velocity difference with respect to the mean of Virgo is less than a velocity dispersion of two ($d_{\rm M87} < R_{200}$, $|v_{\rm gal} - v_{\rm Virgo}| < 2\sigma_{\rm Virgo}$). Among the filament samples, the overlapping galaxies that belong to the Virgo cluster region ($< R_{200}$) have been excluded from the five structures but they are still included in Virgo reference.

%%%%% subsection 2.2
\subsection{Data}
\label{sec2.2:data}

Our sample galaxies are mainly selected from the SDSS DR12 \citep{sdssdr12}. The ``model" magnitudes of $ugriz$ bands have been used to estimate the colors of the galaxies. As shown in Table~\ref{tab1:sample}, the upper radial velocity limit of the sample is 3200~km~s$^{-1}$. The extinction was computed following \citet{schlegel98}, adopting $A(\lambda)/E(B-V) =$~5.155, 3.793, 2.751, 2.086 and 1.479 for \textit{u, g, r, i,} and $z$ magnitudes.

First, the AllWISE\footnote{\href{https://wise2.ipac.caltech.edu/docs/release/allwise/}{https://wise2.ipac.caltech.edu/docs/release/allwise/}} data \citep{allwise} is suitably used as a general SFR tracer (W3 and W4). It is the reliable census of the old stellar population that is less affected by dust attenuation. Among four WISE bands (centered on 3.4, 4.6, 12.0, and 22.0~$\mu$m), we utilize W3 (12.0~$\mu$m) and W1 (3.4~$\mu$m) magnitudes with signal-to-noise ratio (S/N) $>$ 3 to calculate the colors of the galaxies (see Figure~\ref{fig:fa1_w3_w4}). We cross matched between the SDSS and AllWISE database by searching within 5$\arcsec$ separation, which is a reasonable search radius given the point-spread function size ($\sim$6$\arcsec$) of AllWISE. Secondly, as a tracer of more recent star formation activities, the UV magnitudes are required. We cross matched between the GALEX Unique source catalog \citep[GCAT\footnote{\href{https://archive.stsci.edu/prepds/gcat/}{https://archive.stsci.edu/prepds/gcat/}};][]{gcat} and the SDSS within 5$\arcsec$ separation. Near-ultraviolet (NUV) magnitudes, which are very sensitive to dust attenuation, have been corrected, adopting $A(\lambda)/E(B-V) = 7.24$ \citep{yuan13}.

Together with the stellar properties, the {\HI} gas masses of the galaxies are collected from the ALFALFA ``A-grid" catalog \citep[$\alpha$.100;][]{alfalfa,giovanelli05}\footnote{\href{http://egg.astro.cornell.edu/alfalfa/data/}{http://egg.astro.cornell.edu/alfalfa/data/}}. The sensitivity limit of the ALFALFA survey is down to 10$^{6}~M_{\odot}$ with 3$\sigma$. We have taken note of the upper decl. limits of the survey is to be 35$^{\circ}$, hence Groups~1 and 2 are not fully covered. For these regions, we collected additional data from the the HyperLeda database \citep[][\url{http://leda.univ-lyon1.fr/}]{paturel03}. Although the data consists of the detections from inhomogeneous observations, the ALFALFA and the HyperLeda data generally show good agreements as we confirmed using the galaxies available in both datasets. Hence, adding the HyperLeda {\HI} detections does not affect the robustness of our analysis and results.

Stellar masses of the galaxies were estimated by using the SDSS ($g-i$) color \citep{zibetti09}, assuming a universal IMF as described by \citet{chabrier03}. The stellar mass distributions of galaxies in the individual structures are shown in the right panel of Figure~\ref{fig:f2_histo}. In this work, we have applied the lowest mass cut of $M_*=10^7M_{\odot}$ to our sample, based on the stellar mass range of the ALFALFA~$\alpha$.40-SDSS-GALEX data by \citet{huang12}. This choice allows us to retain as many {\HI} detections as possible while remaining within the mass range well-explained by global scaling relations. Note that the $M_*=10^7M_{\odot}$ cut could be lower than the stellar mass range of $M_* \sim 10^8$--$10^{10}~M_{\odot}$, which is commonly probed by optically selected samples in dense environments. We confirm that there is no significant incompleteness in depth across the different catalogs used. Nevertheless, caution is required when interpreting the colors of low-mass galaxies and evaluating the impact of environmental processes. In our samples, we find a relatively high fraction of low-mass galaxies in Filaments~1$-$3, with more than 20\% having stellar masses below $10^{8}M_{\odot}$. This implies that our sample contains a good number of targets that are more vulnerable to the surroundings, and hence very suitable to trace the evolution of galaxies in low-density environments of the local Universe. 
%%%%%%%%%%%%%%%%%%%%%%%%%%%%%%%%%%%

%%%%%%%%%%%%%%%%%%%%%%%%%%%%%%%%%%% Section 3. SF indicators
\section{Multiwavelength Colors as Star Formation Indicators}
\label{sec3:sf_indicators}

We utilize three different multiwavelength color indicators, W3$-$W1, NUV$-r$ and \textit{g$-$r} to trace the star formation properties of the galaxies in the low-density environments. These colors complement each other by capturing different phases of star formation history and minimizing the impact of incomplete data for faint galaxies in certain wavelength ranges. 
\vspace{0.2cm}

\begin{itemize}
    \item \textit{W3$-$W1.} This infrared color is selected to trace the general star formation properties of the galaxies because it is sensitive to strong dust and PAH emission over a timescale of one to a few Gyr in star formation history. A total of 251 galaxies for the five structures are plotted, along with 109 galaxies as Virgo reference. The numbers for each structure are consistent with the available data in each wavelength, as shown in~Table~\ref{tab1:sample}. A detailed comparison between using W3 (12$\mu$m) and W4 (22$\mu$m) is provided in Appendix~B.
    
    \item \textit{NUV$-r$.} This is a direct tracer of recent dust-unobscured star formation, typically occurring within a few hundred Myrs \citep{salim05}. The strong correlation between observed NUV$-r$ color and the {\HI} gas fraction suggests that galaxies exhibiting blue NUV$-r$ colors are gas-dominated \citep[XGASS;][]{catinella18}. For NUV$-r$, 448 galaxies are available from the five structures, with 194 galaxies as Virgo reference.
    
    \item \textit{g$-$r.} This is a widely used optical color for distinguishing between star-forming and quiescent galaxy populations. Among the three different colors, the SDSS \textit{g$-$r} color provides the largest sample. In total, 510 galaxies are selected from the five structures, along with 222 galaxies used as Virgo reference.
\end{itemize} 

%%%%%%%%%%%%%%%%%%%%%%%%%%%%%%%%%%%

%%%%%%%%%%%%%%%%%%%%%%%%%%%%%%%%%%% Section 4. SF maps
\section{Star Formation in Low-density Environments}
\label{sec4:sf_environments}

%%%%%% subsection 4.1
\subsection{Mapping the Star Formation Properties of the Galaxies}

\begin{figure*}
\epsscale{1.15}
\plotone{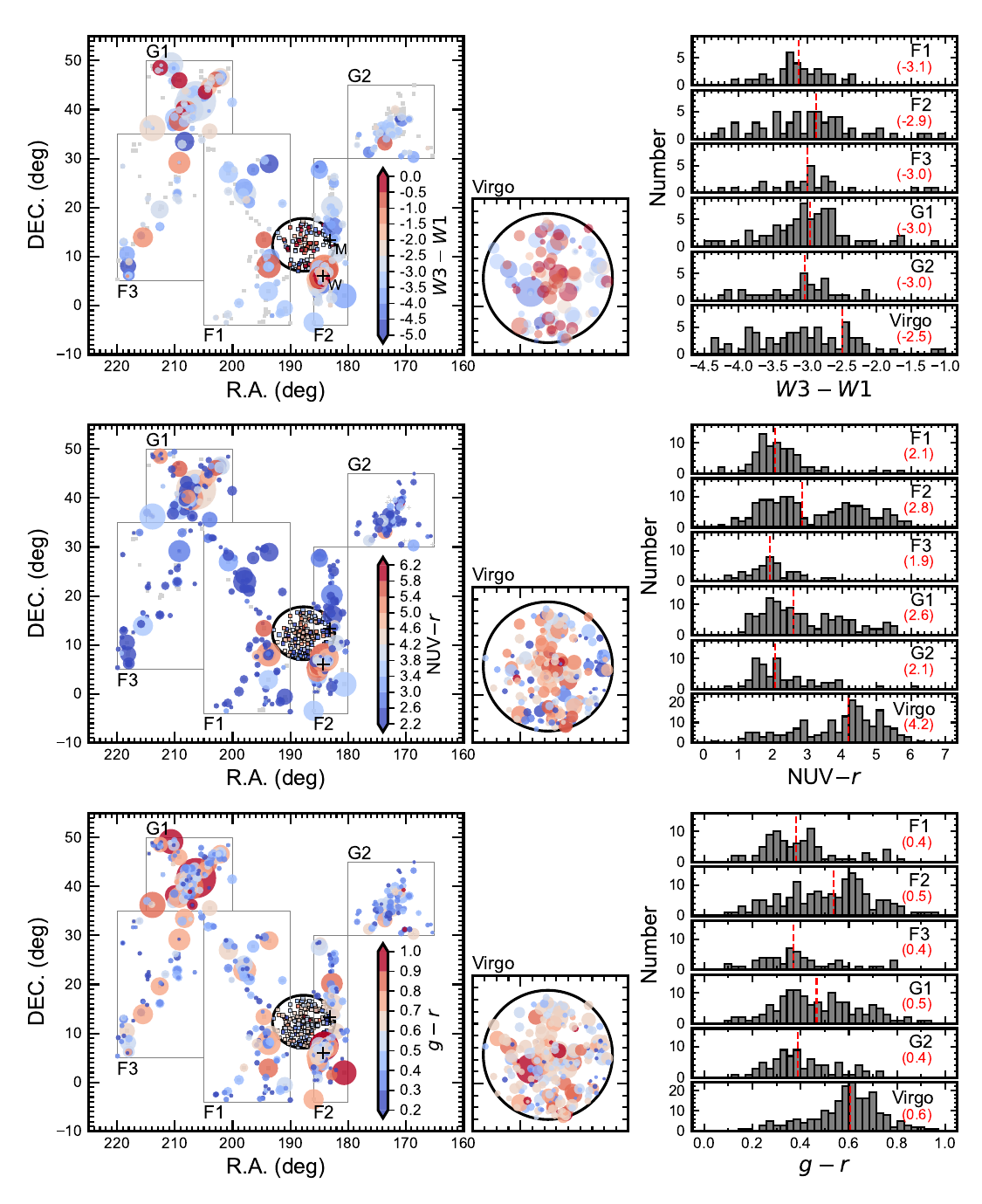}
\caption{Left: Spatial distribution of the galaxies are mapped with three different colors. The color scheme of the symbol indicates W3$-$W1 (top), NUV$-r$ (middle), and \textit{g$-$r} (bottom) colors, respectively. The size of the filled circles shows the stellar mass of the galaxies. Small gray dots in the background present the optically selected members in each structure. The black solid circle in the middle shows $R_{200}$ of Virgo ($\sim$5.4 deg). In $R_{200}$, Virgo reference galaxies are shown as small squares with an identical symbol size. This area is zoomed-in on the right. Right: Number distribution of the sample across the colors. The red dotted line indicates the median of the distribution, with the corresponding values written in red. 
\label{fig:f3_sf_maps}}
\end{figure*}

To investigate how star-forming activities of galaxies are affected in low-density environments, we probed three multiwavelength colors of the galaxies in the selected large-scale structures. Figure~\ref{fig:f3_sf_maps} presents the spatial distribution of the galaxies, traced by different SFR indicators, W3$-$W1, NUV$-r$, and \textit{g$-$r} colors. We aim to probe the star formation quenching or enhancement of galaxies in the low-density environments using the location of red or blue galaxies. 

First, the W3$-$W1 color distribution shows that most galaxies are located in blue sequence region with W3$-$W1 $<$ $-2.5$ (Figure~\ref{fig:f3_sf_maps}; top). The Virgo sample has a median W3$-$W1 color of $-2.5$, making this a reasonable threshold for identifying red galaxies. However, we confirm that some galaxies near the cluster in Filaments~1 and 2, but still outside of $R_{200}$ of Virgo, tend to have red colors with W3$-$W1 $>$ $-2.5$ (14\% and 31\%, respectively). Filament~2 is associated with the W and M~clouds of Virgo \citep{binggeli93,gavazzi99,kim16}, two prominent substructures located on the far side of the cluster, at roughly twice the distance of the cluster center. The presence of red galaxies in this region suggests preprocessing, infalling from the back toward the cluster core. Additionally, there are a few notably blue galaxies with W3$-$W1 $<$ $-4$, suggesting enhanced star formation in the filaments.

Second, the distribution with NUV$-r$ colors show a clearer separation between red and blue galaxies (Figure~\ref{fig:f3_sf_maps}; middle). The observed projected distribution of red galaxies suggests that galaxies with quenched star formation are located in the outskirts of Virgo and Group~1. Among the red galaxies, there are various sizes of the symbol, indicating that the stellar mass range of passively evolving galaxies spreads widely. As seen in the NUV$-r$ histogram, the color bimodality is more noticeable in Filament~2 and Group~1, where a higher fraction of red galaxies than in the other structures are found (28\% in Filament~2 and 16\% in Group~1).

Finally, the distribution of \textit{g$-$r} colors exhibits less clear separation compared to that of the other two colors (Figure~\ref{fig:f3_sf_maps}; bottom). However, we also find some red populations in the same overdense regions in Filaments~1 and 2 as seen in the W3$-$W1 and NUV$-r$ distributions (37\% in Filament~2 and 26\% in Group~1). In the histogram, Filament~2 and Group~1 again show a higher fraction of red populations with their medians located at the redder side than that of any other structures. 

\begin{figure*}
\epsscale{1.2}
\plotone{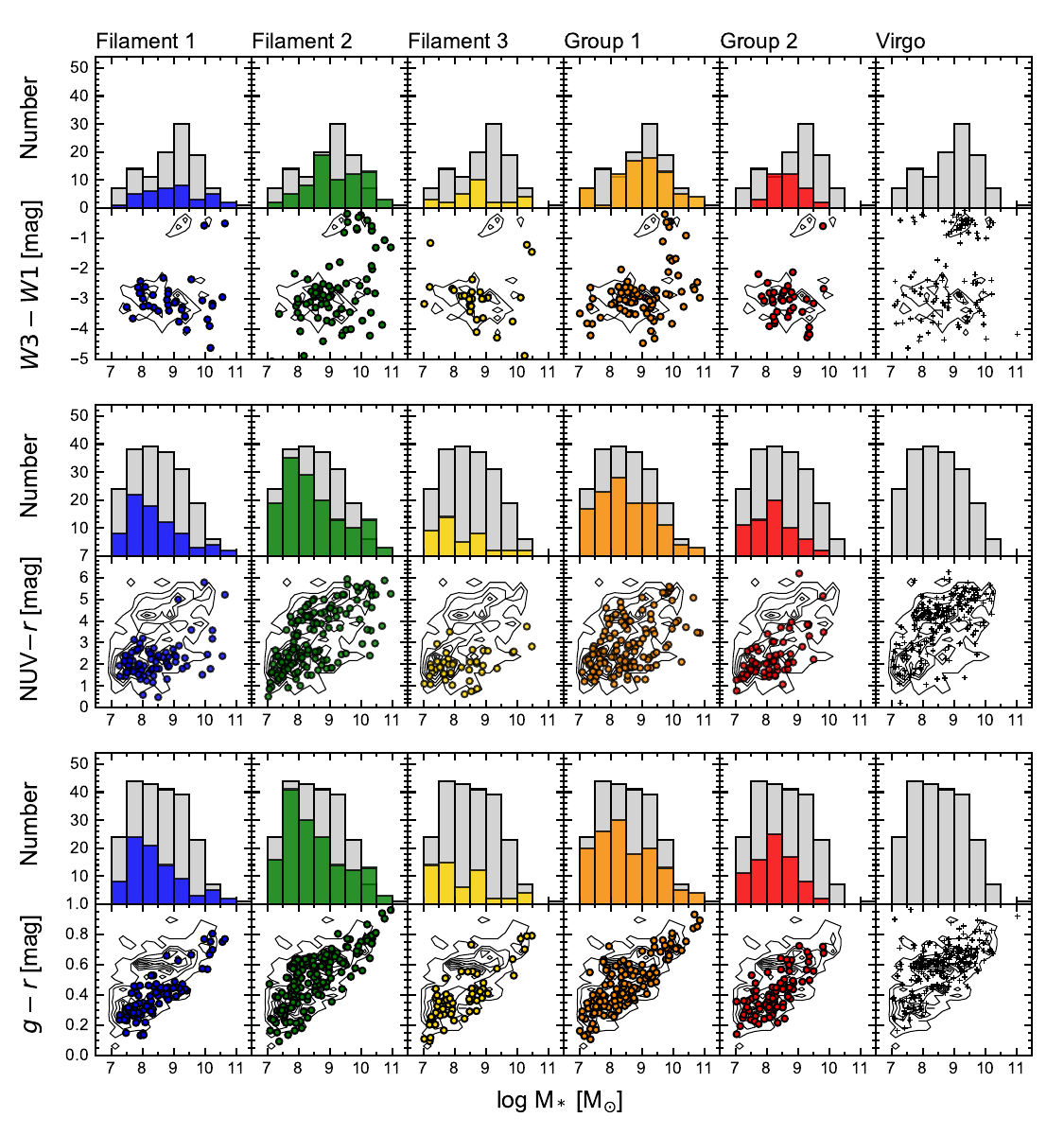}
\caption{Star formation tracer vs. stellar mass relation of galaxies in each structure. The top panel shows W3$-$W1 color, the middle panel shows NUV$-r$ color, and the bottom panel shows \textit{g$-$r} color of the galaxies. Filled circles indicate the galaxies in each structure, and black crosses are Virgo reference sample. The stellar mass distribution is shown at the top of each panel, with gray-shaded bars indicating Virgo reference and colored bars presenting the distribution of the galaxies in each structure.
\label{fig:f4_sf_indicator}}
\end{figure*}

Figure~\ref{fig:f4_sf_indicator} illustrates the color distribution of the galaxies across different stellar mass bins, with corresponding quantities listed in Table~\ref{tab2:sfcolor}. The top of the Figure~\ref{fig:f4_sf_indicator} shows W3$-$W1 colors of the galaxies. The background contour represents the number density of all galaxies within 3$R_{200}$ of Virgo in the range of $cz < 3200$~km~s$^{-1}$. Compared to the background, the five structures exhibit a clearer separation between the red sequence and blue cloud. Many galaxies in the color-stellar mass plane are located in the blue cloud region, whereas Filament~2 and Group~1 have $\sim$10~galaxies in the red sequence region the same as in the Virgo reference sample (red crosses). Additionally, we find a few galaxies with 9.5 $<$ log~$M_{*}$ $<$ 11) in transition between the blue and red sequences, located at $-2$ $\le$ W3$-$W1 $<$ $-1$.

%%%%%%%%%%%%%%%%%%%%%%%%%%%%%%%%%%% Figure 3 (page width) 
%%%%%%%%%%%%%%%%%%%%%%%%%%%%%%%%%%% 

%%%%%%%%%%%%%%%%%%%%%%%%%%%%%%%%%%% Figure 4 (page width) 
%%%%%%%%%%%%%%%%%%%%%%%%%%%%%%%%%%%

%%%%%%%%%%%%%%%%%%%%%%%%%%%%%%%%%%% Table 2 (page width)
\begin{deluxetable*}{lrrrr}
\tablecaption{Stellar Mass and Star Formation Tracer Properties of Galaxies in the Five Large-scale Structures and Virgo 
\label{tab2:sfcolor}}
\setlength{\tabcolsep}{20pt} 
\tablehead{
\colhead{Structure} & \colhead{Log~$M_{*}$\tablenotemark{\scriptsize a}} & \colhead{W3$-$W1} & \colhead{NUV$-r$} & \colhead{\textit{g$-$r}} \\
\colhead{} & \colhead{($M_\odot$)} & \colhead{(mag)} & \colhead{(mag)} & \colhead{(mag)}
 }
\decimalcolnumbers
\startdata
Filament~1 & 7.3 $\sim$ 10.6 (8.4)& $-$4.6 $\sim$ $-$0.5 ($-$3.1) & 0.5 $\sim$ 5.8 (2.1) & 0.1 $\sim$ 0.8 (0.4)\\
Filament~2 & 7.1 $\sim$ 11.0 (8.5) & $-$5.1 $\sim$ $-$0.2 ($-$2.9) & 0.5 $\sim$ 5.9 (2.8) & $-$0.1 $\sim$ 1.0 (0.5)\\
Filament~3 & 7.0 $\sim$ 10.5 (8.2) & $-$5.1 $\sim$ $-$1.2 ($-$3.0) & 0.6 $\sim$ 3.7 (1.9) & 0.1 $\sim$ 0.8 (0.4)\\
\hline
Group~1 & 7.0 $\sim$ 11.5 (8.5) & $-$4.5 $\sim$ $-$0.2 ($-$3.0) & 1.0 $\sim$ 5.6 (2.1) & 0.1 $\sim$ 1.0 (0.5)\\
Group~2 & 7.0 $\sim$ 9.8 (8.2) & $-$5.7 $\sim$ $-$0.6 ($-$3.0) & 0.7 $\sim$ 6.2 (2.1) & 0.1 $\sim$ 0.7 (0.4)\\
\hline
Virgo & 7.0 $\sim$ 11.0 (8.5) & $-$5.3 $\sim$ $-$0.1 ($-$2.5) & 0.2 $\sim$ 6.3 (4.2) & 0.2 -- 1.2 (0.6)\\
\enddata
\tablecomments{~Column~(1): Name of each structure. Column~(2): Logarithmic stellar mass. Columns~(3)$-$(5): W3$-$W1 color, NUV$-r$, and \textit{g$-$r} range, respectively. The numbers in parentheses in Columns~(2)$-$(5) represent their median values.\\
\tablenotemark{\scriptsize a}Stellar masses are from the optical sample, the largest dataset in this work (Column~(5) in Table~\ref{tab1:sample}).}
\end{deluxetable*}
%%%%%%%%%%%%%%%%%%%%%%%%%%%%%%%%%%%

%%%%%%%%%%%%%%%%%%%%%%%%%%%%%%%%%%% Table 3 (page width)
\begin{deluxetable*}{lrrrrrr}
\tablecaption{Number of Red Galaxies within the Five Large-scale Structures and Virgo
\label{tab3:red_num}}
\setlength{\tabcolsep}{6pt} 
\tablehead{
\colhead{Structure} & \multicolumn{6}{c}{Number Fraction of Red Galaxies} \\
\hline
\colhead{} & \multicolumn{2}{c}{W3$-$W1 $>$ $-2.5$} & \multicolumn{2}{c}{NUV$-r > 4.2$} & \multicolumn{2}{c}{\textit{g$-$r} $>$ 0.6} \\
\colhead{} & \colhead{log~$M_* < 8.5$} & \colhead{log~$M_* > 8.5$} & \colhead{log~$M_* < 8.5$} & \colhead{log~$M_* > 8.5$} & \colhead{log~$M_* < 8.5$} & \colhead{log~$M_* > 8.5$}
}
\decimalcolnumbers
\startdata
Filament~1 & 1/12 (8\%$\pm$8\%) & 4/25 (16\%$\pm$7\%) & 0/48 (0\%$\pm$0\%) & 3/29 (10\%$\pm$6\%) & 0/53 (0\%$\pm$0\%) & 10/33 (30\%$\pm$8\%)\\
Filament~2 & 2/15 (13\%$\pm$9\%) & 20/57 (35\%$\pm$6\%) & 11/83 (13\%$\pm$4\%) & 29/59 (49\%$\pm$7\%) & 15/87 (17\%$\pm$4\%) & 41/66 (62\%$\pm$6\%)\\
Filament~3 & 1/10 (10\%$\pm$9\%) & 3/18 (17\%$\pm$9\%) & 0/28 (0\%$\pm$0\%) & 0/14 (0\%$\pm$0\%) & 0/35 (0\%$\pm$0\%) & 6/20 (30\%$\pm$10\%)\\
\hline
Group~1 & 1/20 (5\%$\pm$5\%) & 14/58 (24\%$\pm$6\%) & 6/68 (9\%$\pm$1\%) & 14/57 (25\%$\pm$3\%) & 6/76 (8\%$\pm$2\%) & 29/61 (48$\pm$4\%)\\
Group~2 & 2/15 (13\%$\pm$9\%) & 2/21 (10\%$\pm$6\%) & 0/44 (0\%$\pm$0\%) & 2/18 (11\%$\pm$7\%) & 3/52 (6\%$\pm$1\%) & 8/27 (30$\pm$4\%)\\
\hline
Virgo & 11/32 (34\%$\pm$8\%) & 44/77 (57\%$\pm$6\%) & 33/101 (33\%$\pm$5\%) & 64/93 (69\%$\pm$5\%) & 41/111 (37\%$\pm$5\%) & 82/111 (74$\pm$4\%)\\
\enddata
\tablecomments{~Column~(1): Name of the structure. Columns~(2)$-$(6): Number of red galaxies meeting criteria W3$-$W1 $>$ $-2.5$, NUV$-r > 4.2$, and \textit{g$-$r} $>$ 0.6, relative to the total number of galaxies. These fractions are estimated separately for two different stellar mass groups determined by the median stellar mass of Virgo sample ($< R_{200}$), which is  log~$M_*=8.5$. The uncertainties are derived from the binomial errors, $\sigma = \sqrt{f(1 - f)/N}$.}
\end{deluxetable*}
%%%%%%%%%%%%%%%%%%%%%%%%%%%%%%%%%%% 

The middle row of Figure~\ref{fig:f4_sf_indicator} presents the distribution of NUV$-r$ color as a function of the stellar mass.  In all the structures, we commonly find that a large number of galaxies are in the blue cloud region (NUV$-r < 3$). However, in the relation with NUV$-r$ color which directly measure the recent sSFR, we find that more passive galaxies are located in Filament~2 and Group~1 than that of the W3$-$W1 distribution. The galaxies with suppressed star formation (NUV$-r > 4$) are found in all stellar mass bins above $M_*=10^{7.5}~M_{\odot}$.

At a fixed stellar mass, the galaxies in Filament~2 and Group~1 are found in the wide range of NUV$-r$ colors. Star-forming activities mainly depend on both the mass of the galaxies and the environment; however, this broad mass distribution of red galaxies suggests that the low-density environments may also play an important role in changing the colors of the galaxies.

In the bottom panel of Figure~\ref{fig:f4_sf_indicator} We find that the overall distribution of \textit{g$-$r} color is similar to that of W3$-$W1 or NUV$-r$. However, the blue and red galaxies are mixed smoothly together compared to the discrete distribution in NUV$-r$. Blue galaxies are easily found in all the structures, and Filaments~1 and 3 mostly consist of the blue populations. Only the small number of galaxies are located in the red sequence region (\textit{g$-$r} $>$ 0.6), and they tend to be massive. In Filament~2 and Group~1, a greater fraction of galaxies is found in the red sequence region, such as the Virgo reference sample (background contours and red crosses).

\begin{figure}
\epsscale{1.1}
\plotone{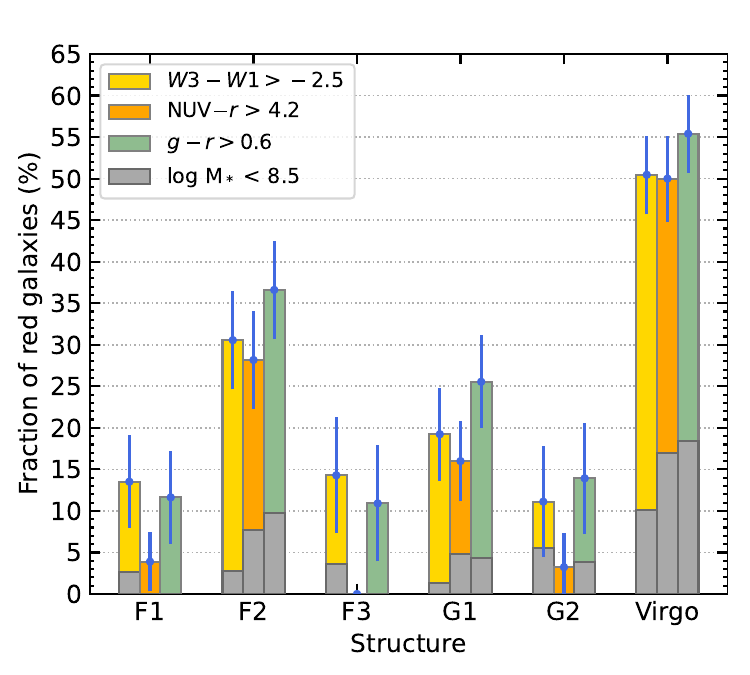}
\caption{Fraction of red galaxies in the sample of large-scale structures. Each color bar presents three different star formation indicators, W3$-$W1 (yellow), NUV$-r$, and \textit{g$-$r} (orange) colors, respectively. The $y$-axis shows the percentage of red galaxies to the total number of galaxies embedded in each region, as summarized in Table~\ref{tab3:red_num}. The red fraction of Virgo reference sample is added for comparison. The gray shade overlaid on the individual bars presents the fraction of low-mass galaxies (log~$M_{*} < 8.5$). The blue error bars indicate the binomial uncertainty for each fraction.
\label{fig:f5_red_fraction}}
\end{figure}

Table~\ref{tab3:red_num} and Figure~\ref{fig:f5_red_fraction} present the fraction of red galaxies in each structure. In all distributions, we find that blue galaxies are dominant in all the structures, but some far more blue or red galaxies are found in each region. The presence of non-negligible low-mass red galaxies in the filaments and groups suggests that the environment may already play a role in quenching star formation, even in low-density regions. However, it is not seen clearly that there is any gradient in color from low- to high-density regions along the filaments.

%%%%%% subsection 4.2
\subsection{Red Populations in the Filaments and Groups}

Filaments~1 and 2 are the most noticeable structures around the Virgo cluster, and hence provide suitable targeting fields to statistically probe the overall color of galaxies outside of the cluster. In Filament~1, we find that a small number of galaxies are located in the red sequence regions of three colors. Five galaxies in W3$-$W1 $>$ $-2.5$, three galaxies in NUV$-r>4.2$, and 10 galaxies in \textit{g$-$r} $>$ 0.6 are likely to evolve passively beyond $R_{200}$ of Virgo in the projected sky. In the bottom panel, much more red populations (22, 40, and 56 galaxies by each color) are found beyond $R_{200}$ of Virgo in Filament~2. Among the entire sample in this filament, 16\%--25\% of galaxies are red, which are mostly located at between 1 and 2 $R_{200}$ of Virgo. Notably, a non-negligible fraction (13\%--17\%) of low-mass red galaxies with log~$M_* = 8.5$ are found in Filament~2. Figure~\ref{fig:fb1_sf_d87_fila1/2} in Appendix~C presents the distribution of three different color indicators in Filaments~1 (top) and 2 (bottom) as a function of the projected distance from the center of Virgo, M87.

We estimate the interquartile range (IQR) ratio (IQR$_{\rm structure}$/IQR$_{\rm Virgo}$) by calculating the IQR, defined as the difference between the third quartile (Q3) and the first quartile (Q1), and comparing it to Virgo. If the color distribution of a structure is similar to Virgo, the ratio should be close to 1. For NUV$-r$, Filament~2 and Group~1 have IQR ratios near 1 (1.24 and 1.01, respectively), whereas other structures range from 0.36 to 0.59. This suggests that their color distributions of Filament~2 and Group~1 are statistically similar to Virgo. For W3$-$W1 and \textit{g$-$r}, the ratio difference is less pronounced than in NUV$-r$, as expected from the less clear separation in Figure~\ref{fig:f3_sf_maps}. However, the results still support the presence of cluster-like populations in Filament 2. Additionally, we performed the Kolmogorov$-$Smirnov (K-S) tests, which confirm that Filament 2 and Group 1 have relatively higher significant showing similar distributions to Virgo. However, due to differences in sample sizes and the discrete nature of the distributions, the absolute significance ($p$-value) is not particularly meaningful because all values fall below the significance threshold, $p=0.05$.

%%%%%%%%%%%%%%%%%%%%%%%%%%%%%%%%%%% Figure 5 (column width) 
%%%%%%%%%%%%%%%%%%%%%%%%%%%%%%%%%%% 
%%%%%%%%%%%%%%%%%%%%%%%%%%%%%%%%%%%

%%%%%%%%%%%%%%%%%%%%%%%%%%%%%%%%%%% Section 5. HI maps
\section{{\HI} Gas in Low-density Environments}
\label{sec5:hi_environments}

%%%%%% subsection 5.1
\subsection{Mapping the {\HI} Gas Properties of the Galaxies}
\label{sec5.1:hi_map}

%%%%%%%%%%%%%%%%%%%%%%%%%%%%%%%%%%% Figure 6 (page width) 
\begin{figure*}
\epsscale{1.2}
\plotone{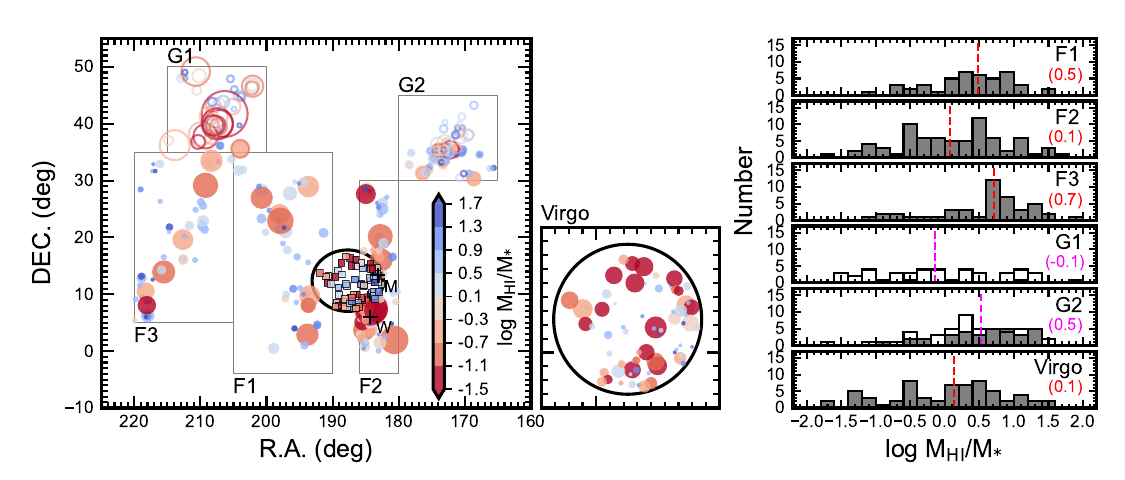}
\caption{Left: Distribution of the {\HI}-to-stellar mass ratio of the galaxies. The color scheme of the symbol indicates the logarithmic {\HI}-to-stellar mass ratio. The {\HI} galaxies in Groups~1 and 2, added from HyperLeda, are shown as open circles. The black solid circle in the middle shows $R_{200}$ of Virgo, and Virgo reference is presented in squares with a single size. A zoomed-in panel for Virgo galaxies is added on the right. Right: Histogram of the {\HI}-to-stellar mass ratio in each structure. The vertical dashed line indicates the median of the distribution, with the corresponding values written. For Groups~1 and 2, the galaxies added from HyperLeda are shown as open bars in the background. The median lines and values in magenta are estimated using the combination of the ALFALFA and HyperLeda samples.
\label{fig:f6_hi_map}}
\end{figure*}
%%%%%%%%%%%%%%%%%%%%%%%%%%%%%%%%%%% 

In addition to the star formation properties of the galaxies in the previous section, we investigate the {\HI} gas-to-stellar mass fraction of the galaxies because {\HI} is the initial fuel reservoir from which molecular gas and, subsequently, stars form. In addition, {\HI} is the most diffuse and abundant interstellar gas, and hence it is an ideal tool to diagnose externally driven impact on galaxies.

Figure~\ref{fig:f6_hi_map} presents the spatial distribution of the galaxies with the logarithmic {\HI}-to-stellar mass ratio in the selected structures. Notably, {\HI}-poor galaxies are prominent in Filament~2; however, a mixture of {\HI}-poor and {\HI}-rich galaxies is observed in all five structures. The separation between these populations is less distinct that what is seen in the NUV$-r$ color distribution in Figure~\ref{fig:f3_sf_maps}.

Due to the decl. limit of the ALFALFA survey ($\sim$36 deg), we have added the {\HI} content of the galaxies in Groups~1 and 2 from the HyperLeda database (open circle). Although they were taken from inhomogeneous observations, the {\HI} mass measurements from the HyperLeda and ALFALFA for galaxies with decl. $<$ $~36^{\circ}$ show good agreement, supporting that our results are statistically robust.

%%%%%%%%%%%%%%%%%%%%%%%%%%%%%%%%%%% Table 4 (page width)
\begin{deluxetable*}{lrrrr}
\tablecaption{Number of Galaxies with Different {\HI} Content in the Five Large-scale Structures and Virgo  
\label{tab4:hi_num}}
\setlength{\tabcolsep}{15pt} 
\tablehead{
\colhead{Structure}  &  \colhead{Total} & \colhead{{\HI}-poor} &  \colhead{{\HI}-poorish} &\colhead{{\HI} Normal-to-Rich}\\
\nocolhead{}  &  \colhead{log~$M_{\rm H~I}/M_*$ Range} & \colhead{} & \colhead{}& \colhead{}
}
\decimalcolnumbers
\startdata
Filament~1 & $-1.1$ $\sim$ 1.6 (0.5) & 0/45 (0\%$\pm$0\%) & 19/45 (42\%$\pm$7\%) &26/45 (58\%$\pm$7\%)\\
Filament~2 & $-2.3$ $\sim$ 1.7 (0.1) & 15/76 (20\%$\pm$5\%) & 37/76 (49\%$\pm$6\%) & 24/76 (32\%$\pm$5\%)\\
Filament~3 & $-1.1$ $\sim$ 1.6 (0.5) & 1/45 (2\% $\pm$ 2\%) & 16/45 (36\%$\pm$7\%) & 28/45 (62\%$\pm$7\%)\\
\hline
Group~1 & $-2.1$ $\sim$ 1.8 ($-0.1$) & 1/46 (2\%$\pm$2\%) & 9/46 (20\%$\pm$6\%) & 36/46 (78\%$\pm$6\%)\\
Group~2 & $-1.8$ $\sim$ 1.6 (0.5) & 1/85 (1\%$\pm$1\%) & 42/85 (49\%$\pm$5\%) & 42/85 (49\%$\pm$5\%)\\
\hline
Virgo & $-2.4$ $\sim$ 1.5 (0.1) & 24/69 (35\%$\pm$6\%) & 31/69 (45\%$\pm$6\%) & 20/69 (36\%$\pm$5\%)\\
\enddata 
\tablecomments{~Column~(1): Name of the structure; Column~(2): Range of {\HI}-to-stellar mass ratio number with the median values; Columns~(3)$-$(5): Number fraction of galaxies in the {\HI}-poor, {\HI}-poorish, and {\HI}-normal-to-rich group classified following the criterion by \citet{mun21}. The definition is mentioned in Section~\ref{sec5.1:hi_map}. The uncertainties are calculated using the binomial errors, as described in the note in Table~\ref{tab3:red_num}.}
\end{deluxetable*}
%%%%%%%%%%%%%%%%%%%%%%%%%%%%%%%%%%% 

\begin{figure}[h]
\epsscale{1.1}
\plotone{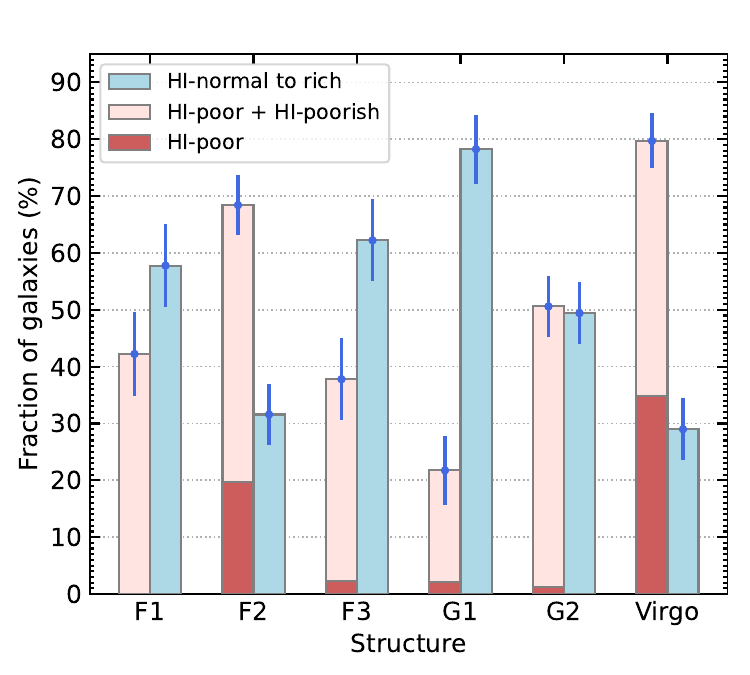}
\caption{Fraction of gas-rich and gas-poor galaxies within the sample large-scale structures. The pink bar represents the fraction of {\HI}-poor and {\HI}-poorish galaxies as detailed in Table~\ref{tab4:hi_num}. {\HI}-poor galaxies are highlighted in red. The blue bars indicate the fraction of {\HI} normal-to-rich galaxies. The blue error represents the binomial uncertainty for each fraction.
\label{fig:f7_hi_fraction}}
\end{figure}

Table~\ref{tab4:hi_num} and Figure~\ref{fig:f7_hi_fraction} present the fraction of galaxies in ``{\HI}-poor'', ``{\HI}-poorish'', and ``{\HI}-normal-to-rich'' groups, classified following the criterion by \citet{mun21}. Our galaxies reside in low-density environments where gas stripping is less active compared to cluster galaxies, resulting in less distinct {\HI} deficiency boundaries \citep{yoon17}. Instead of {\HI} deficiency, \citet{mun21}'s classification utilizes the relation between the stellar mass and gas fraction of Virgo galaxies, considering the deviation from the main sequence relation. Each group fulfills these relations; (i) {\HI}-poor: log~$M_{\rm H~I}/M_* + 0.915~$log~$M_* < (8.478-1.179)$; (ii) {\HI}-poorish: $(8.478-1.179) \leq $~log~$M_{\rm H~I}/M_* + 0.915~$log~$M_* < (8.478-0.42)$; (iii) {\HI}-normal-to-rich: log~$M_{\rm H~I}/M_* + 0.915~$log$M_* \leq (8.478-0.42)$. The binomial error for gas-rich or gas-poor is not significant, with a maximum range of up to 7\%. According to the classification, all the five large-scale structures have a high fraction of {\HI}-normal-to-rich galaxies, but Filament~2 and Group~2 shows a relatively lower fraction (32\% and 49\%, respectively). This suggests that gas stripping in the filament or group environment could be more common than conventionally believed. Within $R_{200}$ of Virgo, there are not many gas-rich galaxies, implying that the gas has already been stripped. 

We find that Filament~2 has a high number of {\HI}-poor galaxies (20\%) compared to any of the other structures. The difference in gas fraction among the filaments is clearly seen from the histogram on the right panel in Figure~\ref{fig:f6_hi_map} and Table~\ref{tab4:hi_num}. In Filament~2, the distribution of the gas fraction appears to hint at two peaks, but the bimodality is not less obvious as seen in the NUV$-r$ color distribution. Filament~2 shows a more extended distribution toward the gas-poor side (negative in $y$-axis), and its median log~$(M_{\rm H~I}/M_{*})$ is $\sim$0.4 to 0.6~dex lower than that of the other two filaments. This is consistent with what we see from Figure~\ref{fig:f3_sf_maps}, showing that Filament~2 has a higher fraction of red galaxies in NUV$-r$. Group~1 galaxies also show low median gas fraction ($-0.1$) as we can expect from the high fraction of red galaxies. Group~1 appears to have fewer {\HI}-poor or {\HI}-poorish galaxies, even though it shows the low {\HI}-to-stellar mass. We find that the {\HI}-detected galaxies in Group~1 tend to have higher stellar mass, and the low $M_{\rm H~I}/M_*$ are mostly from high stellar mass galaxies. This implies that this group genuinely contains massive galaxies or a small amount of {\HI} may not be detectable due to the less uniform collection of {\HI} data in this area. These galaxies tend to fall into the {\HI} normal-to-rich group due to their high stellar masses. The gas-poor galaxies in Filament~2 and Group~1 are located in the same overdense region where the red galaxies are found in Figure~\ref{fig:f3_sf_maps}. This is suggestive that both the stellar and the gas properties of galaxies in this area are influenced by the environments.

%%%%%%%%%%%%%%%%%%%%%%%%%%%%%%%%%%% Figure 7 (column width) 
%%%%%%%%%%%%%%%%%%%%%%%%%%%%%%%%%%% 

%%%%%% subsection 5.2
\subsection{Gas-poor Galaxies in the Filaments and Groups}

We found that a significant number of low SFR galaxies reside outside $R_{200}$ of Virgo. Filament~2 has 15\% of {\HI}-poor galaxies, and 42\% and 49\% {\HI}-poorish galaxies are found in Filaments~1 and 2, respectively. This implies that the total gas budget does not seem to be affected as strongly as the SFR in the low-density environments. The {\HI}-to-stellar mass ratio of the galaxies in Filaments~1 and 2 as a function of the distance from M87 can be seen from Figure~\ref{fig:fb2_hi_d87_fila1/2}. Filament~2 shows more red, gas-deficient and massive galaxies than in Filament~1. With the given sensitivity limits of the datasets we use, we do not find any evident sign of gas deficiency in these structures. Even though NUV$-r$ has a strong correlation with M$_{\rm H~I}$/M$_{*}$, there are very few {\HI}-detected galaxies with gas fractions below 10$\%$. Meanwhile, it seems that the effect of stellar mass is quite apparent since the galaxies with the low gas fraction have a larger circle. However, we note that the stellar mass of these galaxies is average or slightly higher. This suggests that the gas content depends on not only the mass but also the environmental effect. 

As observed in the three star formation colors, the IQR tests show that Filament~2 and Group~1 have IQR ratios close to 1 (0.80 and 1.13, respectively), while other structures range from 0.50 to 0.67. We also performed the K-S test using the {\HI}-to-stellar mass ratio for each structure compared to Virgo. The results indicate that the gas fractions in Filaments 1, 3, and Group 2 are significantly different from that of Virgo. Filament~3 shows the most pronounced difference ($p = 1.74 \times 10^{-5}$), suggesting weaker or different environmental effects. Similarly, Filament~1 ($p = 0.02$) and Group~2 ($p = 0.003$) also show statistically significant differences, though to a lesser extent. In contrast, Filament~2 ($p = 0.43$) and Group~1 ($p = 0.82$) exhibit exceed or are close to the significance threshold of $ p= 0.05$, implying that their gas fraction distributions are statistically similar to that of Virgo. This suggests that some galaxies in these environments may have undergone similar evolutionary processes as those in Virgo, potentially due to comparable gas stripping or accretion mechanisms.

\begin{figure*}[ht]
\epsscale{1.15}
\plotone{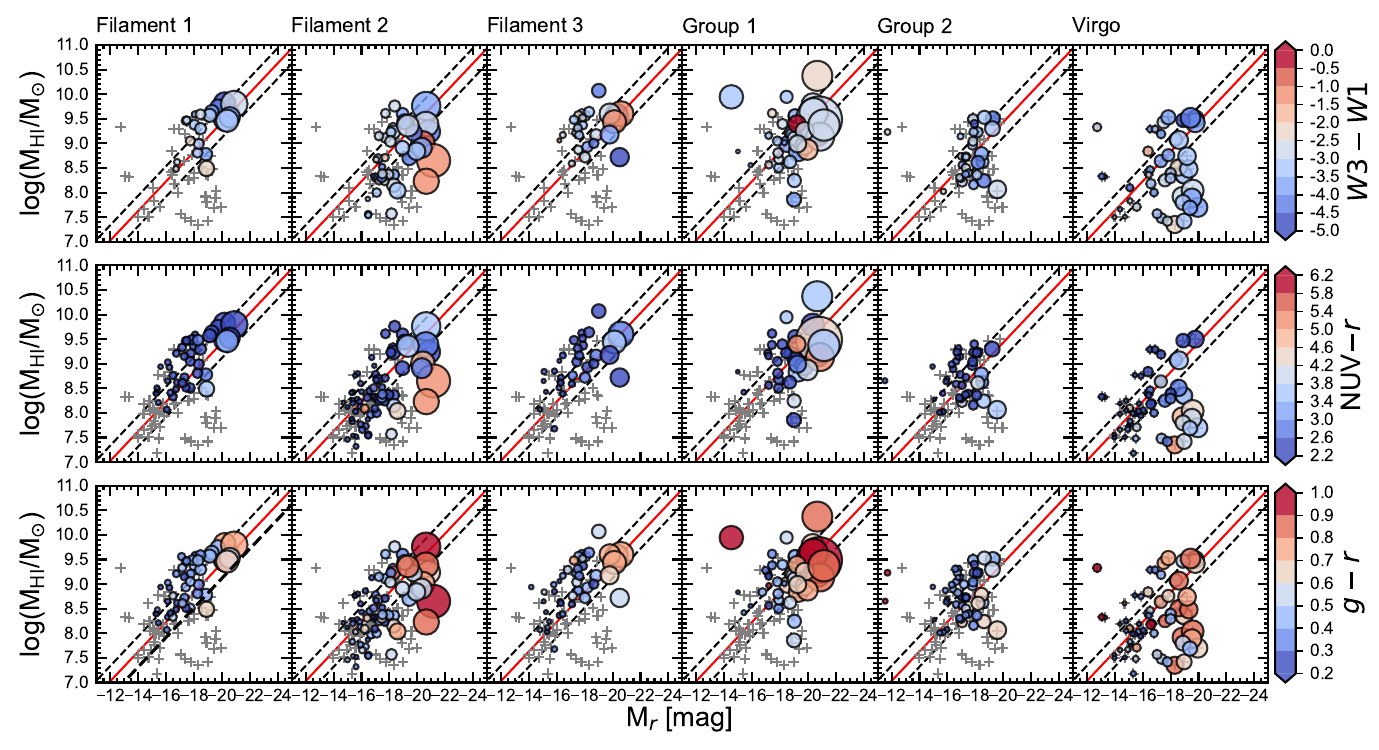}
\caption{{\HI} mass vs. absolute $r$-band magnitude of the galaxies in each region and Virgo. The colors of the symbol indicate W3$-$W1 (top), NUV$-r$ (middle), and \textit{g$-$r} (bottom), respectively. The size of the symbol presents the stellar mass of the sample. The red solid line is the {\HI} scaling relation obtained from the galaxies in NOIRCAT \citep{wong06,wong09} by \citet{denes14}. The dashed lines indicate 1$\sigma$ deviation from the red line. The background gray crosses are Virgo reference galaxies.
\label{fig:f8_scaling_hi_mass}}
\end{figure*}

In Figure~\ref{fig:f8_scaling_hi_mass}, the scaling relation between the {\HI} mass and the absolute magnitude in the $r$-band is presented. We also compared the distribution of our sample to the scaling relation determined by \citet{denes14} (red solid line), which includes all galaxies in NOIRCAT \citep{wong06,wong09} above 2$^{\circ}$ in decl. We find that $\sim$35\% galaxies in Filament~2 deviate by more than 1$\sigma$ from the scaling relation and lean toward the low {\HI} mass side, implying that their gas content is lower than the average of isolated galaxies. The distribution of {\HI} mass of the galaxies is very similar to the wide range observed in the Virgo reference sample. In Filament~2, the low-mass galaxies tend to have normal-to-higher {\HI} gas content in general, while a few gas-poor galaxies have low stellar mass. Besides, at a given stellar mass, the range of gas content and star formation properties are widely spread. This suggests that not only massive red galaxies but also passively evolving red galaxies that are less massive, and found in low-density regions, are suggestive of the environmental impact. 

In addition, in Group~1, there are many red galaxies, but many of them have a relatively normal or slightly higher gas content. This may suggest that the net change of the {\HI} gas mass in these galaxies is not as significant as the global color change of the stellar population. It can also suggest that there is an offset in time and evolutionary stage between changes in the gas reservoir and an observable change in star formation history. This might suggest additional gas supply. This will be discussed in the following section.

%%%%%%%%%%%%%%%%%%%%%%%%%%%%%%%%%%% Figure 8 (page width) 
%%%%%%%%%%%%%%%%%%%%%%%%%%%%%%%%%%% 

%%%%%%%%%%%%%%%%%%%%%%%%%%%%%%%%%%% Figure 9 (page width) 
\begin{figure*}
\epsscale{1.15}
\plotone{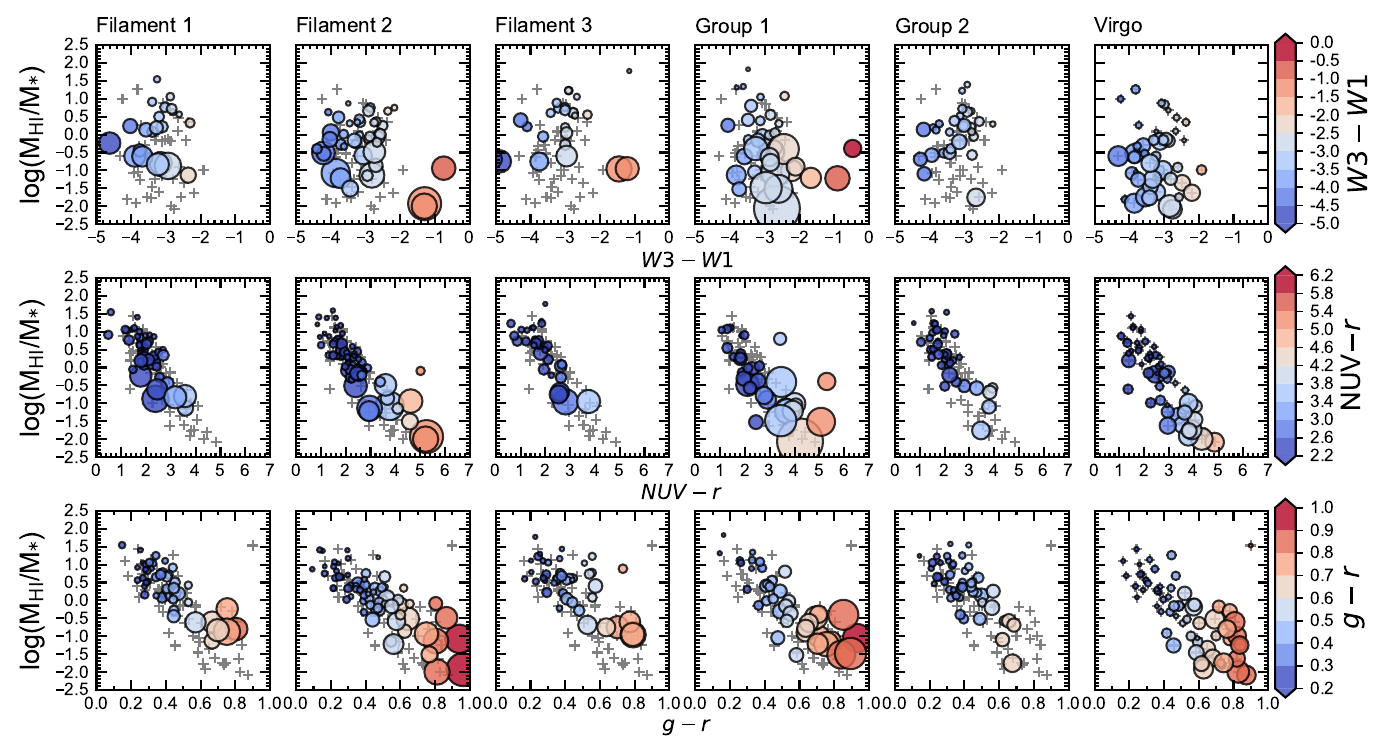}
\caption{{\HI} gas fraction as a function of the stellar mass of the galaxies in each region and Virgo. The colors of the symbol show W3$-$W1 (top), NUV$-r$ (middle), and \textit{g$-$r} (bottom), respectively. The size of the symbol indicates the stellar mass of sample galaxies. The background gray crosses are Virgo reference galaxies.
\label{fig:f9_scaling_gas_fraction}}
\end{figure*}
%%%%%%%%%%%%%%%%%%%%%%%%%%%%%%%%%%% 
%%%%%%%%%%%%%%%%%%%%%%%%%%%%%%%%%%%

%%%%%%%%%%%%%%%%%%%%%%%%%%%%%%%%%%% Section 6. Discussion
\section{Discussion}
\label{sec6:discussion}

%%%%%% subsection 6.1
\subsection{Physical Processes on Galaxies in Filaments and Groups}

\textit{Red Galaxies and SF Quenching (Mass versus Environment).} The color distribution in Figure~\ref{fig:f3_sf_maps} shows that most galaxies generally show normal or enhanced star formation, similar to normal galaxies in isolated environments. But at the same time a non-negligible number of galaxies have low SFR, similar to the galaxies inside the cluster. Even though the details of how star formation is quenched in these low-density environments is not exactly understood, mass and environment are known to be the main drivers of the suppression of star formation \citep[e.g.,][]{peng12}. In our sample, some galaxies with low SFR are massive. In this case, it is hard to separate the impact of mass from the impact of the environment on the star formation. However, in our sample, the fraction of low-mass galaxies is high, as shown in Figure~\ref{fig:f2_histo}, and the degree of star-forming activities is diverse at a given stellar mass. In particular, the low-mass red galaxies (log~$M_*$ $<$ 8.5) are mainly found in Filament~2 and Group~1. To date, the evolution of red dwarf galaxies in low-density environments is not well understood, and those in Filament~2 and Group~1 can be potentially good targets to probe the role of the environment on the quenching of star formation in low-mass galaxies. We posit that local environmental effects are likely to impact the evolution of low-mass galaxies, in concert with secular evolutionary processes.

\textit{Preprocessing and Other Studies.} The red galaxies that we find in the filaments and groups lead us to consider the presence of preprocessing \citep{fujita04} of galaxies in low-density environments. Indeed, the signs of preprocessing have been recognized by many previous studies. First, stellar and gas components can be disturbed by neighboring galaxies in the filaments or groups \citep{hess13}. Galaxies in low-density environments have a greater chance to encounter other galaxies than cluster members due to the slow speed \citep{mihos04}. Host halo masses or dynamics of groups can also directly affect the quenching of star formation in galaxies. \citet{lee22} provide observational evidence supporting the preprocessing within loosely bound galaxy groups through CO~imaging. \citet{brown23} find a reduction in the molecular gas densities, even in the unperturbed Virgo galaxies, which is tentatively attributed to preprocessing before cluster infall. Second, cold gas in galaxies can be influenced by the surrounding medium \citep{gunn72}. Interactions between the interstellar gas of a galaxy and the surrounding intergalactic medium can be one potential reason for causing the removal of gas in galaxies. Although it is generally not easy to observationally trace the intergalactic gas mostly due to its relatively low density and temperature, X-ray detections have been made \citep{mulchaey03} in the overdense regions of Filament~2 (NGC~4168 and NGC~4261 groups), Groups~1 (HCG~68) and 2 (NGC~3665 group). Based on the color and {\HI} distribution as shown in Figures~\ref{fig:f3_sf_maps} and \ref{fig:f6_hi_map}, we did not observe any clear gradient in SFR or {\HI} fraction along the two filaments around the Virgo cluster. This implies either that there might be no filament-scale environmental effects on galaxy evolution or that there are many overlapping processes that are averaging-out various effects in the filaments.

\textit{Where does the Preprocessing Occur?} Based on Figure~\ref{fig:f3_sf_maps}, we mentioned a few passively evolving galaxies found in Filaments~1 and 2 outside $R_{200}$ of Virgo. \citet{guo17} show that the quenching effect of star formation can start beyond one virial radius of clusters. \citet{zinger18} present that the significant removal of the star-forming disk is only within 0.5~virial radius, but the disk is potentially preprocessed outside one virial radius. Since Virgo is less virialized, local substructures within or outside $R_{200}$ can contribute toward changes in galaxies' stellar and gaseous properties.

\textit{Optical Morphology.} We examined the morphological types of red galaxies in Filaments 1 and 2, using the Extended Virgo Cluster Catalog (EVCC) \citep{kim14}, which provides detailed morphological classifications. Among red galaxies with $g-r > 0.6$, 29\% (two out of seven galaxies) in Filament 1 and 61\% (53 out of 87 galaxies) in Filament 2 are early types (dwarf ellipticals, ellipticals, or lenticulars). The presence of numerous early-type galaxies and a notable fraction of disky or irregular red galaxies in low-density environments is intriguing. We find no clear trend between stellar mass and morphology, suggesting that environmental effects may be as influential as stellar mass in shaping galaxy properties. A more detailed investigation would be valuable for future studies.
 
\textit{Backsplashing of Galaxies.} If Filaments~1 and 2 are physically related to Virgo, ``\textit{backsplashing}" \citep{gill05} might be another possible way to interpret the physical processes on the red galaxies outside $R_{200}$ of the cluster. This suggests that the galaxies already passed the core of the cluster in the past, and moved out to 1.5$-$2$R_{200}$ in the orbits. \citet{mun21} investigated a potential backsplashing population, identified as ``{\HI}-poor'' (see the note in~Table~\ref{tab4:hi_num}) beyond the virial radius ($|\Delta v/\sigma_{cl} \leq 1$ and $R_{\rm proj}/R_{200} \geq 1.5$). Their findings indicate that these galaxies may have experienced mild gas loss attributed to ram pressure stripping in the past, though they do not rule out the possibility of preprocessing. The backsplashed populations are known to have low velocities with respect to the mean velocity of clusters and low gas content \citep[e.g.,][]{yoon17,jaffe15}. In our sample, the low gas content of a few galaxies is in agreement with that of the backsplashed galaxies, whereas the radial velocities of the sample are quite high compared to the mean of Virgo. Thus, for these cases, the star formation is more likely to have been suppressed due to the local environments rather than the cluster core-crossing a while ago.

%%%%%% subsection 6.2
\subsection{Timescales of SF Quenching and Gas Depletion}

In previous sections, we have used three different star formation indicators, and each tracer samples a different number of galaxies, as seen in Table~\ref{tab1:sample}. It follows that each panel in Figure~\ref{fig:f3_sf_maps} appears similar, though the color distributions are not exactly identical. For instance, W3$-$W1 can be used as a general SFR indicator ($\sim$100~Myr$-$1~Gyr), whereas NUV$-r$ reveals the recent star formation history from young stars (a few hundred Myrs). In general, \textit{g$-$r} probes a longer timescale compared to NUV$-r$ ($\sim$1$-$10~Gyr). The W3$-$W1 color is an indicator of recent star forming activities rather than a more general star formation history of a longer timescale as it traces emissions from PAH molecules that reside in star-forming {H\kern 0.2em\sc ii} regions. Also, WISE selects only bright galaxies due to the surface brightness limitations, thus using W3$-$W1 color alone might not be sufficient to interpret the star formation history of the faint sample. However, the overall stellar mass distribution from the WISE data appears uniform (Figure~\ref{fig:f4_sf_indicator}), and there is no clear discontinuity in the number histogram (Figure~\ref{fig:f3_sf_maps}). Therefore, the lack of high S/N data for faint galaxies in the W3 band is unlikely to contribute to a less distinct trend in W3$-$W1. Although both NUV$-r$ and \textit{g$-$r} colors use $r$-band magnitudes, NUV and $g$-band magnitudes lead us to probe different timescales of star formation. NUV$-r$ color is a powerful tool for tracing recent star formations, whereas \textit{g$-$r} color probes a longer timescale of star formation. Compared to the SFR as a function of clustocentric distance (Figure~\ref{fig:fb1_sf_d87_fila1/2}), the {\HI} gas fraction (Figure~\ref{fig:fb2_hi_d87_fila1/2}) less clearly indicates the impact of environments. This may suggest that our SFR tracers may span too broad a timescale, limiting us to observe these changes with sufficient time resolution. However, the effect of stellar mass is quite evident, as shown by many larger circles representing galaxies with low {\HI}-to-stellar mass fractions, as studied in previous works, such as \citet{huang12}. Also, the number of galaxies with low {\HI} gas fractions is lower than that of the red galaxies outside $R_{200}$ of Virgo.

Figure~\ref{fig:f8_scaling_hi_mass} shows a few galaxies which are below the {\HI} scaling relation but still blue in all colors. They are found in Filament~2, Groups~1 and 2, suggesting that gas is depleted earlier than the quenching of star formation. Hence, these galaxies have already been quenched in star formation and in the final episodes of star formation prior to complete cessation. In contrast, in Figure~\ref{fig:f8_scaling_hi_mass}, a few massive galaxies with low star formation (\textit{g$-$r} $>$ 0.6) are lying on or even above the {\HI} scaling relation. Figure~\ref{fig:f9_scaling_gas_fraction} also shows the presence of red yet gas-rich galaxies. For instance, at a given color in Filament~2, \textit{g$-$r} $\sim$ 0.8, the range of gas fraction on the $y$-axis spans approximately $\sim$2.0~dex, indicating the difference in gas fraction between gas-rich and gas-poor galaxies is about 100~times. These red and gas-rich galaxies are found in the similar regions in both the W3$-$W1 and NUV$-r$ plots. Not only Filament~2 but also Group~1 and Filament~3 exhibit a few red and gas-rich galaxies. Compared to the W3$-$W1 and \textit{g$-$r} colors, the NUV$-r$ plot shows an overall tighter correlation with fewer outliers. NUV$-r$ is a direct tracer of the recent star formation history of young stars in galaxies. This may suggest that the red galaxies with normal-to-high {\HI} gas content might accrete more gas from their surroundings after finishing their initial star formation.
%%%%%%%%%%%%%%%%%%%%%%%%%%%%%%%%%%%

%%%%%%%%%%%%%%%%%%%%%%%%%%%%%%%%%%% Section 7. Summary
\section{Summary and Conclusions}
\label{sec7:summary}

We have mapped the star formation properties of the galaxies in the large-scale filaments and groups around the Virgo cluster. We use three different SFR tracers, W3$-$W1, NUV$-r$ and \textit{g$-$r} colors of the sample, and find a few passively-evolving galaxies in the low-density environments. The red galaxies are found outside $R_{200}$ of Virgo on the projected sky, suggesting the preprocessing of the galaxies prior to entering the cluster. In addition, the {\HI} gas fraction of the galaxies has also been mapped to compare between the gas and stellar properties of the galaxies. We conclude that:

\begin{itemize}

\item[(i)] Among the three different SFR tracers we used, W3$-$W1 and \textit{g$-$r} colors show similar trends, tracing timescales of one to a few Gyr in star formation history. NUV$-r$ provides the clearest evidence of the presence of a quenched/passively-evolving population in Filament~2 and Group~1. Our analysis more effectively traces recent star formation activities over the past few hundred Myr to 1~Gyr.

\item[(ii)] No SFR or {\HI} gas fraction gradient is observed along the filaments, suggesting either minimal filament-scale environmental effect on galaxy evolution or the simultaneous influence of multiple processes that obscure any signature. Alternatively, the timescales of our SFR tracers may be too broad to resolve these changes. However, we still find convincing signs of quenched star formation beyond $R_{200}$ of Virgo on the projected sky. 

\item[(iii)] From the {\HI}-to-stellar mass ratio of the galaxies, we find that gas-poor galaxies exist even in the low-density environments. However, unlike the star formation properties, our results suggest that gas fraction is likely to depend on the stellar mass than the environment in these low-density regions. This implies that even if the current SFR from NUV$-r$ and the {\HI} fraction do not show clear gradients toward higher densities, those galaxies that reside in the filaments are already on the way to being quenched or undergoing gas accretion.
 
\end{itemize}

We note that our approach may have potential uncertainties due to projection effects; however, the identified structures are likely gravitationally bound and strong targets for studying physical mechanisms in low-density environments. Together with our results, a high-resolution {\HI} imaging study of galaxies may provide more direct evidence for environmental effects on galaxy evolution in low-density environments. We observed 14~galaxies in the filaments around Virgo in {\HI} to find preprocessing signatures outside the cluster (Yoon et al. in preparation). Combining {\HI} imaging with the global picture from this work, we anticipate being able to enhance our understanding of environmental impacts on galaxies before they dive into denser environments. 
%%%%%%%%%%%%%%%%%%%%%%%%%%%%%%%%%%%

%%%%%%%%%%%%%%%%%%%%%%%%%%%%%%%%%%% Acknowledgements
\begin{acknowledgments}
\label{sec:acknowledgments}

We are indebted to the anonymous referee for the constructive comments and valuable suggestions, which were greatly helpful in improving the manuscript. A.C. acknowledges support by the National Research Foundation of Korea (NRF), No. RS-2022-NR070872 and RS-2022-NR069020. H.Y. acknowledges support by the Endeavour Research Fellowship of the Australian Government (No. 5030\_2016). This work was supported by the Global-LAMP Program of the National Research Foundation of Korea (NRF) grant funded by the Ministry of Education (No. RS-2023-00301976). This work was supported by the National Research Foundation of Korea (NRF) grant funded by the Korea government (MSIT) (RS-2025-00516062). 

We acknowledge the work of the ALFALFA team \citep[\href{http://egg.astro.cornell.edu/alfalfa/}{http://egg.astro.cornell.edu/alfalfa;}][]{giovanelli05,alfalfa} in releasing their useful catalogs. This research has made use of the NASA/IPAC Extragalactic Database \href{https://ned.ipac.caltech.edu/}{(NED; https://ned.ipac.caltech.edu/)}, the HyperLeda database \href{http://leda.univ-lyon1.fr/}{(LEDA; http://leda.univ-lyon1.fr/)}, the Sloan Digital Sky Survey \href{http://www.sdss.org/}{(SDSS; http://www.sdss.org/)}, the AllWISE data \citep{allwise}, the GALEX Unique source catalog \citep[GCAT;][]{gcat}, \verb'Astropy' \citep[\href{http://www.astropy.org/}{http://www.astropy.org/};][]{astropy13,astropy18,astropy22}, \verb'astroML' \citep[\href{https://www.astroml.org/}{https://www.astroml.org/};][]{astroml12}, and \verb'CosmoloPy' package \href{https://roban.github.io/CosmoloPy/}{(https://roban.github.io/CosmoloPy/)}. 

\end{acknowledgments}
%%%%%%%%%%%%%%%%%%%%%%%%%%%%%%%%%%%

%%%%%%%%%%%%%%%%%%%%%%%%%%%%%%%%%%% References

%%%%%%%%%%%%%%%%%%%%%%%%%%%%%%%%%%%

%%%%%%%%%%%%%%%%%%%%%%%%%%%%%%%%%%% Appendix
\newpage
\appendix
\label{sec:appendix}
\restartappendixnumbering

%%%%%%%%%%%%%%%%%%%%%%%%%%%%%%%%%%%  Appendix A
\restartappendixnumbering
\label{sec:appendix_a}
\section{Statistics from a Wider Field Around Virgo}

\begin{figure*}[b]
\gridline{\fig{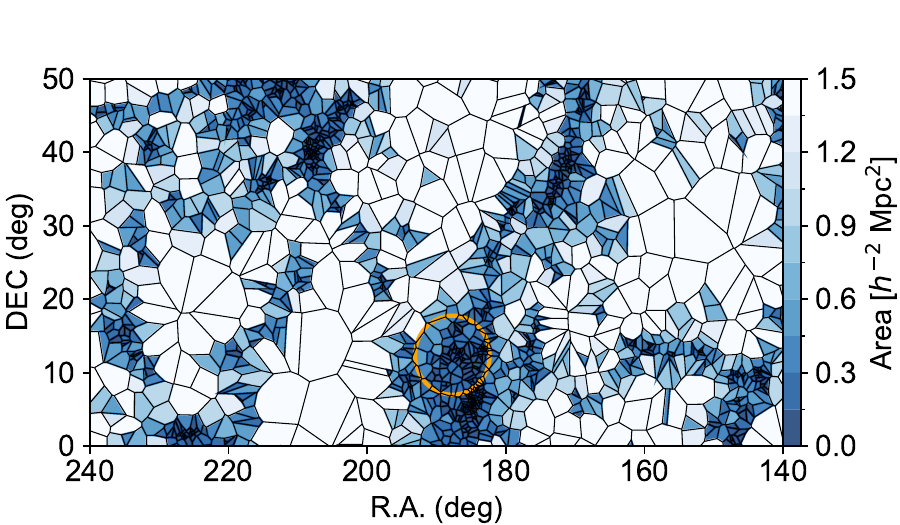}{0.6\textwidth}{}
\fig{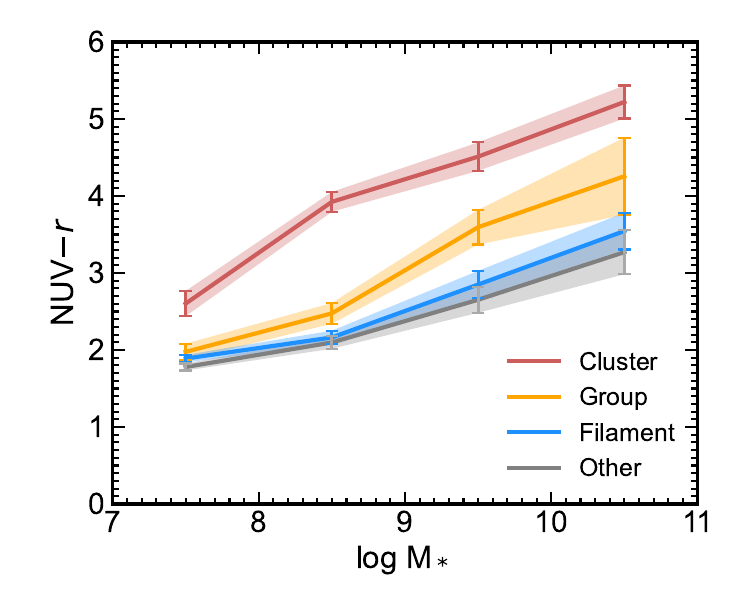}{0.4\textwidth}{}}
\caption{Left: Voronoi tessellation of galaxies in Virgo field with color-coded by the density. Darker cells indicate dense environments. $R_{200}$ of Virgo is shown in the circle in the middle. Right: Stellar mass versus NUV$-r$ of each structure identified by the Voronoi tessellation. ``Other'' environments indicate lower density region than filaments. The shade indicates 1$\sigma$ deviation from the mean of each distribution.
\label{fig:fc1_voronoi}}
\end{figure*}

\begin{figure*}
\gridline{\fig{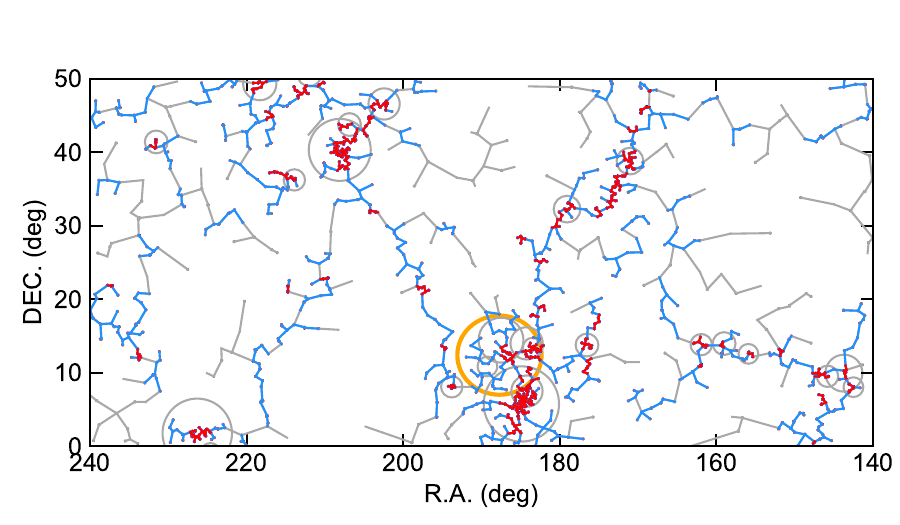}{0.6\textwidth}{}
\fig{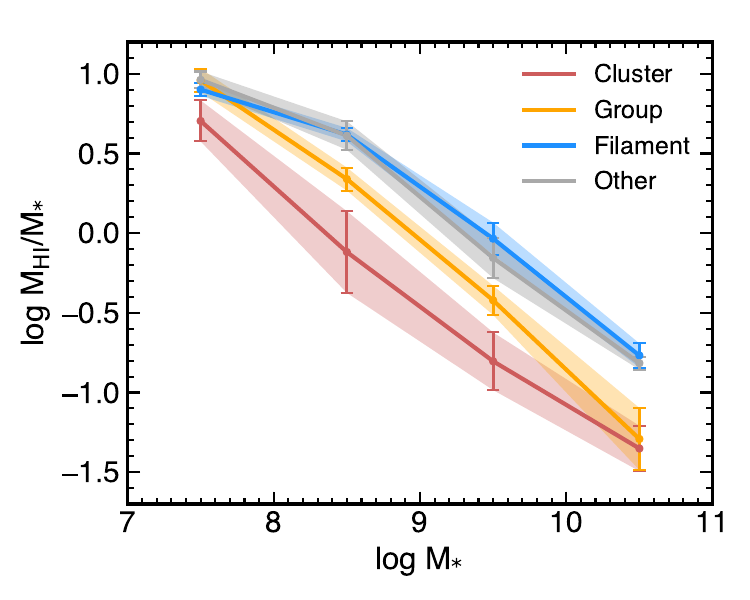}{0.4\textwidth}{}}
\caption{Left: MST of galaxies in Virgo field. Blue branches connect the galaxies in ``Filaments" (within 1~Mpc linking length), and red branches connect the galaxies in ``Groups". Gray lines connect the remaining ``other'' galaxies that do not belong to any specific structure. gray circles present the location of groups ($n_{\rm mem}>5)$ in the field, identified by \citet{tempel12}. $R_{200}$ of Virgo is shown as an orange circle in the middle. Right: {\HI} gas fraction vs. stellar mass of each structure identified by the MST. The meaning of colors and shades are the same as in Figure~\ref{fig:fc1_voronoi}.
\label{fig:fc2_mst}}
\end{figure*}

In this section, we construct a wide-field map around Virgo by using the Voronoi tessellation and the MST methods. Figures~\ref{fig:fc1_voronoi} and \ref{fig:fc2_mst} confirm the identification of our selected five structures and further support the presence of red or gas-poor galaxies in Figures~\ref{fig:f3_sf_maps} and \ref{fig:f6_hi_map} with a larger number of samples.

Figure~\ref{fig:fc1_voronoi} is a Voronoi diagram of the Virgo field, and the size and color of each cell represent the density of the surroundings, i.e. darker and smaller cells indicate denser environments. The sample consists of 1012 galaxies, selected from the optical-UV matches using SDSS and GCAT catalogs as described in Section~\ref{sec2.2:data}. The sky coverage spans R.A. from 140$^{\circ}$ to 240$^{\circ}$ and decl. from 0$^{\circ}$ to 50$^{\circ}$. We set the radial velocity range of the sample from 1800 to 3200~km~s$^{-1}$, which maximizes the inclusion of the five structures while minimizing foreground contamination. The Voronoi method clearly identifies the filamentary structures. We define the lowest density environment as ``other'' if the density is lower than that of the filaments. Based on this identification, we obtain the NUV$-r$ color versus stellar mass plot of galaxies in different density regions. We find that group galaxies show a greater deviation from the ``other'' sample than the filament galaxies, but the deviation is not as significant as that of the cluster sample. However, it is not clearly seen that the colors of low-mass galaxies in filaments are very different from that in ``other''. Despite a considerable fraction of red dwarfs, the difference in NUV$-r$ among the environments becomes smaller as the stellar mass decreases. Nevertheless, a noticeable difference in NUV$-r$ between filament and ``other'' environments is observed for massive galaxies, consistent with the high fraction of red galaxies in Filament~2, as discussed in Section~\ref{sec4:sf_environments}. The results in Figure~\ref{fig:fc1_voronoi} indicate that the red galaxies identified in Filament~2 and Group~1 are statistically unusually red. The mean NUV$-r$ range for filament and ``other" environment is between 1.8 and 3.5 across the stellar mass range, as shown in Figure~\ref{fig:fc1_voronoi}. The red galaxies are predominantly found in the Filament~2 and Group~1, with NUV$-r$ values above 4 in Figure~\ref{fig:f4_sf_indicator}, which are rare in the overall sample. This suggests potential environmental processing occurring in these regions.

%%%%%%%%%%%%%%%%%%%%%%%%%%%%%%%%%%% Figure A1 (page width) 
%%%%%%%%%%%%%%%%%%%%%%%%%%%%%%%%%%% 

%%%%%%%%%%%%%%%%%%%%%%%%%%%%%%%%%%% Figure A2 (page width) 
%%%%%%%%%%%%%%%%%%%%%%%%%%%%%%%%%%% 

Figure~\ref{fig:fc2_mst} shows the MST map of the Virgo field, identified by using the linking lengths for filaments and groups ($\sim$1~Mpc). The sample galaxies are selected in the same sky coverage and velocity range as those of Figure~\ref{fig:fc1_voronoi}. A total of 1523 galaxies from SDSS are used to identify the structures shown in the left panel. Using the MST map, we compare the {\HI} gas fraction among different density environments for 511 galaxies with available {\HI} data, as shown in the right panel. We find that group galaxies exhibit lower gas fractions than those in filament or even lower density environment (``other''). However, the difference in gas fractions between filaments and ``other'' environment is not as significant as the difference in NUV$-r$ throughout the entire stellar mass range. It is possible that some filament galaxies have been dropped from our analysis whereas some galaxies included in the ``other'' sample are rather located at the edge of large-scale structures in different velocity ranges, potentially obscuring subtle differences between the two environments. The overall trend in {\HI} gas is less obvious than NUV$-r$, as seen from the difference between Figures~\ref{fig:f3_sf_maps} and \ref{fig:f6_hi_map}. However, the lower gas fraction found in galaxies with log~$M_* > 8$, compared to those in filaments and ``other'' environments, suggests the presence of gas-poor populations, as in Group~1 in the earlier section.

Our statistics by using the Voronoi and MST methods support that galaxies in groups are much more processed than those in filaments or lower density regions, but less processed than cluster galaxies. However, it is not seen clearly that the color and gas fraction of low-mass galaxies in filaments differ significantly from those in regions with even lower density.
%%%%%%%%%%%%%%%%%%%%%%%%%%%%%%%%%%%

%%%%%%%%%%%%%%%%%%%%%%%%%%%%%%%%%%% Appendix B
\restartappendixnumbering
\label{sec:appendix_b}
\section{Comparison Between WISE W3 and W4 Bands}

\begin{figure}[h!]
\epsscale{0.5}
\plotone{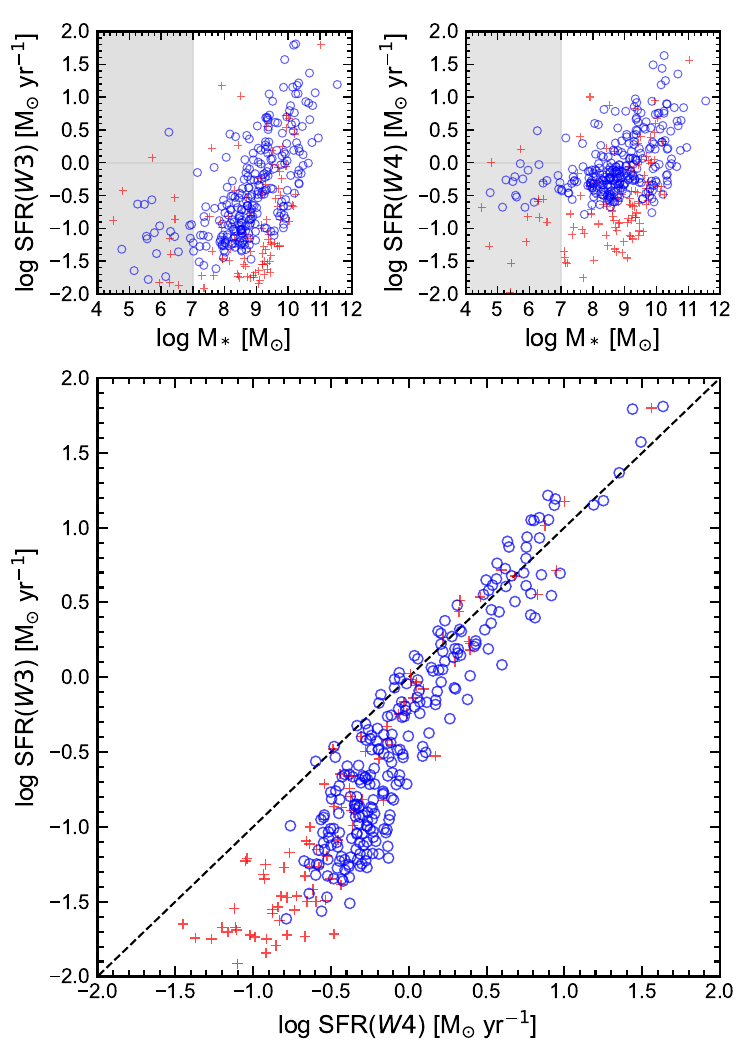}
\caption{Top: SFR, estimated by using W3 (left) and W4 (right) versus stellar mass of the sample with S/N $>$ 3. Open symbols indicate all galaxies belong to the five large-scale structures. Red crosses are Virgo references. Shaded regions include the galaxies below the lowest limit of the stellar mass in this study (log~$M_*=7$). Bottom: Comparison of SFRs based on W4 ($x$-axis) and W3 ($y$-axis). The galaxies with log~$M_*<7$ are excluded. The dashed line indicates a one-to-one line. The symbols are the same as shown in the top panel.
\label{fig:fa1_w3_w4}}
\end{figure}

In Section~\ref{sec3:sf_indicators}, an infrared color, W3$-$W1, is chosen to trace general star formation properties of the galaxies with the timescales of one to a few Gyr in star formation history. Among four WISE bands, the longest band, W4 (22~$\micron$), which originates from the emission reradiated by the dust around star-forming regions, is known to be the most reliable star formation indicator \citep{cluver14}. However, our sample contains a high fraction of low-mass galaxies which mostly have a S/N less than 3 in W4. Thus, we decide to use W3 instead of W4. As verified with the galaxies with S/N $>$ 3 in W4, W3 and W4 fluxes shows a good linear correlation (see Figure~\ref{fig:fa1_w3_w4}) and the use of W3 should not make significant differences in our analysis where relative comparisons across the five structures and the reference are more important. 

In the top panel of Figure~\ref{fig:fa1_w3_w4}, the SFRs estimated by using W3 and W4 \citep{jarrett13} are shown as a function of the stellar mass of the galaxies. Compared to W3, the number of the galaxies with S/N $>$ 3 in the low SFR regions is smaller in W4. In the bottom panel of Figure~\ref{fig:fa1_w3_w4}, the SFRs by W4 and W3 are directly compared. They follow the one-to-one correlation down to $\sim 1~M_{\odot}$~yr$^{-1}$. However, the sensitivity limit of the W4-based SFR is only $\sim 1~M_{\odot}$~yr$^{-1}$, and W3$-$W1 is a better tracer for probing star formation of low-mass galaxies than W4$-$W1.

%%%%%%%%%%%%%%%%%%%%%%%%%%%%%%%%%%% Figure B1
%%%%%%%%%%%%%%%%%%%%%%%%%%%%%%%%%%% 
%%%%%%%%%%%%%%%%%%%%%%%%%%%%%%%%%%%

%%%%%%%%%%%%%%%%%%%%%%%%%%%%%%%%%%% Appendix C
\restartappendixnumbering
\label{sec:appendix_c}
\section{Star Formation and {\HI} Gas Properties versus Distance from M87}

%%%%%%%%%%%%%%%%%%%%%%%%%%%%%%%%%%% Figure C1 (page width) 
\begin{figure*}[hb]
\epsscale{1.2}
\plotone{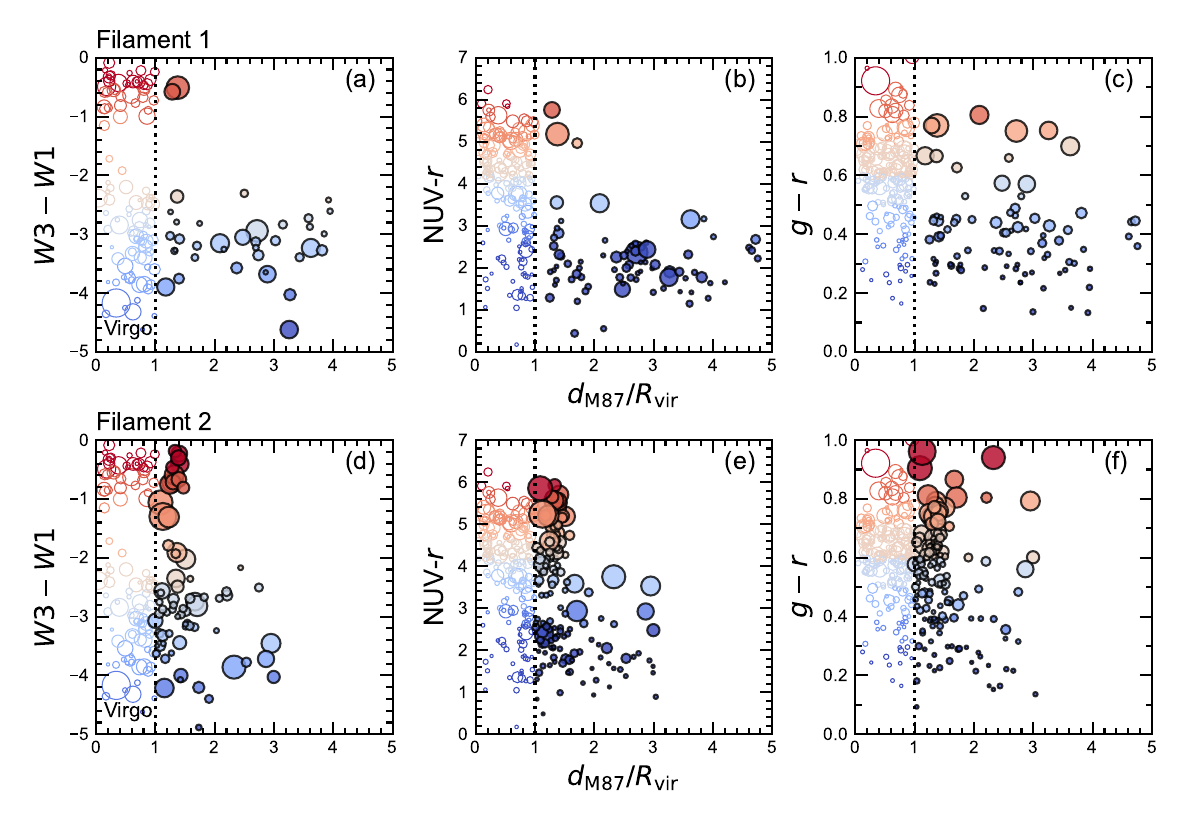}
\caption{Three star formation tracers as a function of the distance from M87 for Filaments~1 (top) and 2 (bottom). The $x$-axis of the panels shows the distance from M87 normalized by $R_{200}$ of Virgo ($\sim$5.4 deg). The $y$-axis indicates W3$-$W1 (left), NUV$-r$ (middle), \textit{g$-$r} (right) colors of the sample, respectively. The color scheme and the size of the symbol are the same as shown in Figure~\ref{fig:f3_sf_maps}. For comparison, the Virgo reference sample is added within $R_{200}$ region (left side of the dotted line) as open circles with the same color scheme.
\label{fig:fb1_sf_d87_fila1/2}}
\end{figure*}
%%%%%%%%%%%%%%%%%%%%%%%%%%%%%%%%%%% 

%%%%%%%%%%%%%%%%%%%%%%%%%%%%%%%%%%% Figure C2 (page width) 
\begin{figure}[ht]
\epsscale{0.5}
\plotone{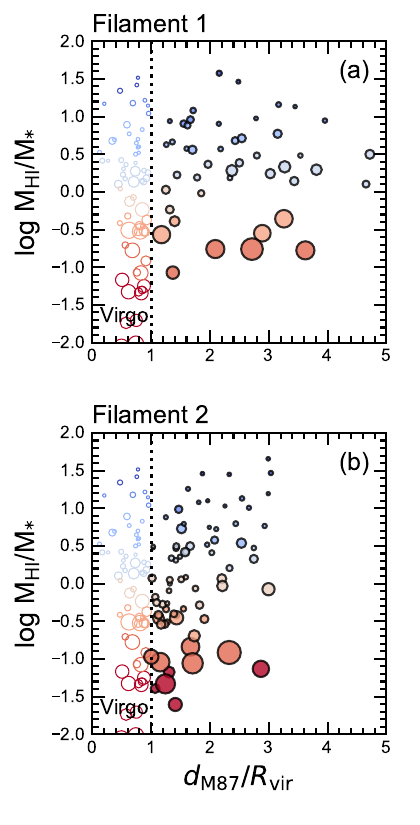}
\caption{{\HI} gas fraction of the galaxies in Filaments~1 (top) and 2 (bottom) as a function of the distance from M87. The colors and sizes of the symbol are the same as shown in Figure~\ref{fig:f6_hi_map}. Virgo reference is presented in open circle. The distance from M87 is normalized by $R_{\rm 200}$ of Virgo ($\sim$5.4 deg).
\label{fig:fb2_hi_d87_fila1/2}}    
\end{figure}
%%%%%%%%%%%%%%%%%%%%%%%%%%%%%%%%%%% 

Figure~\ref{fig:fb1_sf_d87_fila1/2} presents the distribution of three different color indicators in Filaments~1 (top) and 2 (bottom) as a function of the projected distance from the center of Virgo, M87. These two filaments are the most noticeable structures around the Virgo cluster, and hence provide suitable targeting fields to statistically probe the overall color of galaxies outside of the cluster. According to \citet{kim16}, Filament~1 is unlikely to be connected directly to Virgo, and instead it is located at the rear side of the cluster. They also mention that Filament~2 can be a sheet-like structure rather than filaments, together with W and M substructures of Virgo \citep{binggeli93,gavazzi99}. If so, the analysis based on the distance from the center of Virgo in Figure~\ref{fig:fb1_sf_d87_fila1/2} might be debatable. Nonetheless, these filaments are still good representatives of low-density environments and suited to address the evolution of galaxies outside of the dense cluster environment. 

In Filament~1, we find that the small number of galaxies are located in the red sequence regions of three colors. Five galaxies in W3$-$W1 $>$ $-2.5$, three galaxies in NUV$-r>4.2$, and 10 galaxies in \textit{g$-$r} $>$ 0.6 are likely to evolve passively beyond $R_{200}$ of Virgo in the projected sky. In the bottom panel, much more red populations (22, 40, and 56 galaxies by each color) are found beyond $R_{200}$ of Virgo in Filament~2. Among the entire sample in this filament, 16\%--25\% of galaxies are red, which are mostly located at between 1 and 2 $R_{200}$ of Virgo. Notably, a non-negligible fraction (13\%--17\%) of low-mass red galaxies with log~$M_* = 8.5$ are found in Filament~2.

Similar to Figure~\ref{fig:fb1_sf_d87_fila1/2}, Figure~\ref{fig:fb2_hi_d87_fila1/2} presents the {\HI}-to-stellar mass ratio of the galaxies in Filaments~1 and 2 as a function of the distance from M87. In Figure~\ref{fig:fb1_sf_d87_fila1/2}, we found that a significant number of low SFR galaxies reside outside $R_{200}$ of Virgo. Meanwhile, in Figure~\ref{fig:fb2_hi_d87_fila1/2}, only Filament~2 has 15\% of {\HI}-poor galaxies, and 42\% and 49\% {\HI}-poorish galaxies are found in Filaments~1 and 2, respectively. This implies that the total gas budget does not seem to be affected as strongly as the SFR in the low-density environments, as shown in Figure~\ref{fig:fb1_sf_d87_fila1/2}. The gas content is widely spread across all of the distance bins. Filament~2 shows more red, gas-deficient, and massive galaxies than in Filament~1. With the given sensitivity limits of the datasets we use, we do not find any evident sign of gas deficiency in these structures.
%%%%%%%%%%%%%%%%%%%%%%%%%%%%%%%%%%% 

\end{CJK*}
\end{document}